\documentclass[aps,prd,10pt,notitlepage,nofootinbib,superscriptaddress,showkeys,showpacs]{revtex4-1}
\pdfoutput=1
\usepackage{amstext,amsmath,amssymb,amsfonts,bbm,euscript,color,dsfont}
\usepackage[latin1]{inputenc}
\usepackage{epsfig}
\usepackage{hyperref}
\usepackage{amsthm}
\usepackage{subfigure}
\usepackage{color}
\usepackage{multirow}
\usepackage{tikz}

\usetikzlibrary{arrows,backgrounds}

\usepackage{verbatim}

\usepackage{graphicx}

\usepackage{latexsym}

\newcommand{\bea}{\begin{eqnarray}}	
\newcommand{\eea}{\end{eqnarray}}
\newcommand{\be}{\begin{equation}}	
\newcommand{\ee}{\end{equation}}
\newcommand{\beq}{\begin{equation}}	
\newcommand{\eeq}{\end{equation}}

\newcommand{\Z}{{\mathbb Z}}
\newcommand{\C}{{\mathbb C}}

\newcommand{\dd}{{\textrm{d}}}

\def\R{\relax\ifmmode {\mathbb R}  \else${\mathbb R}$\fi}
\def\C{\relax\ifmmode {\mathbb C}  \else${\mathbb C}$\fi}
\def\Z{\relax\ifmmode {\mathbb Z}  \else${\mathbb Z}$\fi}
\def\N{\relax\ifmmode {\mathbb N}  \else${\mathbb N}$\fi}
\def\I{\relax\ifmmode {\mathbb I}  \else${\mathbb I}$\fi}

\begin{document}

\title{A non-perturbative study of matter field propagators in Euclidean Yang-Mills theory in linear covariant, Curci-Ferrari and maximal Abelian gauges}

%\begin{comment}

\author{M.~A.~L.~Capri}\email{caprimarcio@gmail.com}

\affiliation{UERJ $-$ Universidade do Estado do Rio de Janeiro,\\
Departamento de F\'isica Te\'orica, Rua S\~ao Francisco Xavier 524,\\
20550-013, Maracan\~a, Rio de Janeiro, Brasil}

\author{D.~Fiorentini}\email{diegodiorentinia@gmail.com}

\affiliation{UERJ $-$ Universidade do Estado do Rio de Janeiro,\\
Departamento de F\'isica Te\'orica, Rua S\~ao Francisco Xavier 524,\\
20550-013, Maracan\~a, Rio de Janeiro, Brasil}

\author{A.~D.~Pereira}\email{aduarte@if.uff.br}

\affiliation{UERJ $-$ Universidade do Estado do Rio de Janeiro,\\
Departamento de F\'isica Te\'orica, Rua S\~ao Francisco Xavier 524,\\
20550-013, Maracan\~a, Rio de Janeiro, Brasil}

\affiliation{UFF $-$ Universidade Federal Fluminense,\\
Instituto de F\'{\i}sica, Campus da Praia Vermelha,\\
Avenida General Milton Tavares de Souza s/n,\\ 
24210-346, Niter\'oi, RJ, Brasil}

\author{S.~P.~Sorella}\email{silvio.sorella@gmail.com}

\affiliation{UERJ $-$ Universidade do Estado do Rio de Janeiro,\\
Departamento de F\'isica Te\'orica, Rua S\~ao Francisco Xavier 524,\\
20550-013, Maracan\~a, Rio de Janeiro, Brasil}

%\end{comment}

\begin{abstract}
In this work, we study the propagators of matter fields within the framework of the Refined Gribov-Zwanziger theory, which takes into account the effects of the Gribov copies in the gauge-fixing quantization procedure of Yang-Mills theory. In full analogy with the pure gluon sector of the Refined Gribov-Zwanziger action,  a non-local long-range term in the inverse of the Faddeev-Popov operator is added in the matter sector.  Making use of the recent BRST invariant formulation  of the  Gribov-Zwanziger framework achieved in \cite{Capri:2015ixa,Capri:2016aqq,Capri:2015nzw,Pereira:2016fpn,Capri:2016gut}, the propagators of scalar and quark fields in the adjoint and fundamental representations of the gauge group are worked out explicitly in the linear covariant, Curci-Ferrari and maximal Abelian gauges. Whenever lattice data are available, our results exhibit good qualitative agreement.

%\
 
%\noindent Pacs numbers: ...\\  
%\noindent Key words: ... \\ 
%\noindent Report numbers: xxxxx

\end{abstract}

\maketitle

\tableofcontents

%%%%%%%%%%%%%%%%%%%%%%%%%%%%%%%%%%%%%%%%%%%%%%%%%%%%%%%%%%%%%%
\section{Introduction}
%%%%%%%%%%%%%%%%%%%%%%%%%%%%%%%%%%%%%%%%%%%%%%%%%%%%%%%%%%%%%%

While quantum chromodynamics (QCD) is well understood at high energies, where perturbation theory is reliable due to asymptotic freedom, the low energy sector remains a challenging open problem in theoretical Physics. In the infrared region, perturbation theory breaks down and non-perturbative techniques are needed.  A full control of the infrared regime of QCD would provide a fundamental understanding of the confinement of quarks and gluons, a goal not achieved till now.

Different approaches which take into account non-perturbative effects in QCD were devised in the last decades, see \cite{Alkofer:2000wg,Binosi:2009qm,Greensite:2011zz,Brambilla:2014jmp,Deur:2016tte}. Up to now, the interplay of such approaches was able to produce non-trivial results. Though,  a complete consistent picture of the mechanism behind colour confinement is still lacking.

One approach to deal with the confinement problem is the non-perturbative study of the correlation functions of the theory. Functional techniques based on the Dyson-Schwinger equations and on the functional renormalization group as well as numerical lattice simulations have been employed in the analysis of the infrared behavior of the correlation functions. In particular,  the two-point gluon correlation function has been object of very intensive investigations. In fact, the infrared structure of two-point gluon correlation function, e.g. the gluon propagator, turns out to encode important features which are interpreted as signals of confinement. For instance, lattice numerical simulations as well as computations based on the Dyson-Schwinger equations show that the gluon propagator exhibits a violation of the reflection positivity. As such, it  cannot be associated with a physical excitation of the spectrum of the theory. This property is interpreted as a manifestation of confinement, see \cite{vonSmekal:1997ohs,vonSmekal:1997ern,Cucchieri:2004mf,Dudal:2013yva,Cornwall:2013zra}. One has to keep in mind that the gluon propagator is a gauge-dependent quantity. Nevertheless, it still contains important information about such (un)physical elementary fields, being  the simplest correlation function one might compute. In the last decade, the gluon propagator has been studied in great detail in the Landau gauge, due to its special features, namely the transversality of the propagator itself and the important property of having a useful lattice formulation which has allowed for a numerical study of the gluon propagator on large lattices. More precisely, the most recent lattice simulations point towards an infrared suppressed gluon propagator which attains a finite value at zero momentum in four and three space-time dimensions, while it vanishes at zero momentum in two space-time dimensions. One says that in three and four dimensions the gluon propagator is of decoupling/massive type, while in two dimensions it is of scaling type, see \cite{Maas:2011se,Cucchieri:2007rg,Aguilar:2008xm,Fischer:2008uz,Maas:2008ri,Cucchieri:2009zt,Maas:2007uv}.

Besides the aforementioned functional and numerical lattice approaches, an analytical framework which takes into account the existence of the Gribov copies \cite{Gribov:1977wm} occurring in the Faddeev-Popov quantization of gauge theories has received increasing interest in the recent years. The so-called Refined Gribov-Zwanziger setup captures the effects of the spurious gauge copies as well as of additional non-perturbative effects related to the existence of dimension two-condensates, giving rise to an effective infrared action, the Refined Gribov-Zwanziger action, yielding a gluon propagator of the decoupling type which is in very good agreement with the most recent lattice data in both four and three space-time dimensions. In two dimensions, infrared singularities forbid the formation of the dimension two-condensates and the refinement does not take place. As a consequence, the gluon propagator turns out to be of the scaling type.  In this paper, we focus on the Refined Gribov-Zwanziger  formulation. An extensive review of the developments of this framework is presented in Sect.~\ref{overviewRGZ}.

Nonetheless, in QCD, in addition of the pure gluon sector, one has to face also the complex issue of quark confinement, to which different strategies have been devoted, see \cite{Greensite:2011zz,Brambilla:2014jmp}. As far as the Refined Gribov-Zwanziger framewrok is concerned, a possible mechanism to take into account matter confinement was proposed in \cite{Capri:2014bsa,Dudal:2013vha,Palhares:2016wqn,Capri:2016aqq} in the Landau gauge. More precisely, as it will be reviewed  in Sect.~\ref{overviewRGZ}, within the Refined Gribov-Zwanziger approach in the Landau gauge, the non-perturbative effect of the Gribov copies is accounted for by restricting the domain of integration in the functional integral to a certain region $\Omega$, called the Gribov region, which is defined by demanding that the Faddeev-Popov operator $\EuScript{M}(A)$ is strictly positive, so that it is invertible within $\Omega$. Such a restriction enables us to eliminate a large set of copies. In practice, the restriction to $\Omega$ is achieved by adding to the starting Faddeev-Popov action an additional non-local term, known as the horizon function,  which contains the inverse of the operator $\EuScript{M}(A)$. It is precisely the addition of this additional long-range term which is responsible for the infrared modifications of the gluon propagator, which turns out to be a confining propagator, exhibiting complex poles and lacking the K{\"a}ll{\'e}n-Lehmann representation. Remarkably, the non-local horizon function can be cast in local form through the addition of a suitable set of auxiliary fields. The resulting action is multiplicatively renormalizable to all orders.   

The proposal made in  \cite{Capri:2014bsa,Dudal:2013vha,Palhares:2016wqn,Capri:2016aqq} consists in generalizing the introduction of the non-local horizon function to the matter sector, in complete analogy with the gluon sector. Moreover, as in the gluon sector, the non-local matter coupling term can be cast in local form, giving rise to a fully local and renormalizable action.   In  \cite{Capri:2014bsa,Dudal:2013vha,Palhares:2016wqn,Capri:2016aqq}, this prescription was implemented for scalar fields in the adjoint representation of the gauge group and for spinor fields in the fundamental representation. The whole procedure was carried out in the Landau gauge, for which the Refined Gribov-Zwanziger setup was well established, see also  \cite{Capri:2015pfa} and references therein for the maximal Abelian gauge.

Very recently, the Refined Gribov-Zwanziger framework has been extended to the class of the linear covariant and Curci-Ferrari gauges \cite{Capri:2015ixa,Capri:2016aqq,Capri:2015nzw,Pereira:2016fpn,Capri:2016gut}, allowing, in particular, to establish the independence from the gauge parameter of the gauge invariant correlation functions as well of the poles of the transverse part of the gluon propagator. In the light of such developments, it seems natural to ask ourselves how matter fields should be coupled to the Refined Gribov-Zwanziger action in such gauges. This is precisely the aim of the present work. 

The paper is organized is follows: Sect.~\ref{overviewRGZ} contains an overview of the Refined Gribov-Zwanziger action, covering the Landau,  the linear covariant, the Curci-Ferrari as well as the maximal Abelian gauge. After that, in Sect.~\ref{noscalaradjoint},  we describe the non-perturbative coupling of scalar fields in the adjoint representation of the gauge group within the Refined Gribov-Zwanziger action in the aforementioned gauges. Subsequently, in Sect.~\ref{npspinor}, we work out the non-perturbative coupling of quark fields and its consequences on the propagator.  Finally, we collect our conclusions. To keep the paper self-contained as much as possible, we have added two appendices devoted to the details of our construction as well as to the conventions used throughout the paper.

%%%%%%%%%%%%%%%%%%%%%%%%%%%%%%%%%%%%%%%%%%%%%%%%%%%%%%%%%%%%%%
\section{Overview of the non-perturbative BRST invariant formulation of the Refined Gribov-Zwanziger framework}\label{overviewRGZ}
%%%%%%%%%%%%%%%%%%%%%%%%%%%%%%%%%%%%%%%%%%%%%%%%%%%%%%%%%%%%%%
In this section, we review the recently proposed BRST invariant formulation of the Refined Gribov-Zwanziger action in linear covariant \cite{Capri:2015ixa,Capri:2015nzw}, Curci-Ferrari \cite{Pereira:2016fpn} and maximal Abelian gauges \cite{Capri:2015pfa}. For the benefit of the reader, we start with a brief overview of the Gribov problem in the Landau gauge for which the Refined Gribov-Zwanziger action was originally constructed. 

%%%%%%%%%%%%%%%%%%%%%%%%%%%%%%%%%%%%%%%%%%%%%%%%%%%%%%%%%%%%%%
\subsection{The Gribov problem in the Landau gauge} \label{gribovlandau}
%%%%%%%%%%%%%%%%%%%%%%%%%%%%%%%%%%%%%%%%%%%%%%%%%%%%%%%%%%%%%%

Let us consider Yang-Mills theory in $d$ Euclidean dimensions with $SU(N)$ gauge group quantized in the Landau gauge, namely  $\partial_{\mu}A^{a}_{\mu}=0$. The Faddeev-Popov procedure results in the gauge-fixed action:

\begin{equation}
S_{\mathrm{FP}}=\int \dd^dx\left(\frac{1}{4}F^{a}_{\mu\nu}F^{a}_{\mu\nu}+b^{a}\partial_{\mu}A^{a}_{\mu}+\bar{c}^{a}\partial_{\mu}D^{ab}_{\mu}(A)c^{b}\right)\,,
\label{ov1}
\end{equation}
with the field strength $F^{a}_{\mu\nu}$ and the covariant derivative $D^{ab}_{\mu}$ in the adjoint representation of the gauge group given by

\begin{eqnarray}
F^{a}_{\mu\nu}&=&\partial_{\mu}A^{a}_{\nu}-\partial_{\nu}A^{a}_{\mu}+gf^{abc}A^{b}_{\mu}A^{c}_{\nu}\,,\nonumber\\
D^{ab}_{\mu}&=&\delta^{ab}\partial_{\mu}-gf^{abc}A^{c}_{\mu}\,. 
\label{ov2}
\end{eqnarray}
The parameter $g$ stands for the gauge coupling\footnote{The coupling $g$ is dimensionless in $d=4$.} and $f^{abc}$ are the real and totally antisymmetric structure constants of the gauge group. The fields $(\bar{c}^{a}, c^a)$ denote the Faddeev-Popov ghosts, while $b^a$ is the Lagrange multiplier implementing the Landau gauge condition. Nevertheless, as investigated by Gribov in \cite{Gribov:1977wm}, the action \eqref{ov1} is plagued by the existence of gauge copies, {\it i.e.} equivalent gauge configurations which still obey the gauge condition.  This fact can be observed very concretely by considering a gauge field configuration $A^{a}_{\mu}$ which satisfies the Landau gauge condition and another gauge field $A'^{a}_{\mu}$ which is connected to $A^{a}_{\mu}$ by an infinitesimal gauge transformation:

\begin{equation}
A'^{a}_{\mu}=A^{a}_{\mu}-D^{ab}_{\mu}\xi^b\,,
\label{ov3}
\end{equation} 
with $\xi^a$ being the infinitesimal gauge parameter of the transformation. If the Landau gauge condition were ideal, \textit{i.e.} if it were selecting only  one representative $A^a_\mu$ per gauge orbit\footnote{A gauge orbit of a given configuration $A^a_\mu$ is the set of all gauge fields related to $A^{a}_{\mu}$ by a gauge transformation.}, then $A'^{a}_{\mu}$ would not obey anymore the Landau gauge condition, $\partial_{\mu}A'^{a}_{\mu}\neq 0$. Therefore, from eq.\eqref{ov3} we should have

\begin{equation}
\partial_{\mu}A'^{a}_{\mu}=\partial_\mu A^{a}_{\mu} -\partial_{\mu}D^{ab}_{\mu}\xi^b = -\partial_{\mu}D^{ab}_{\mu}\xi^b \neq 0\,,
\label{ov4}
\end{equation} 
where the condition $\partial_{\mu}A^{a}_{\mu}= 0$ was employed. Hence, eq.\eqref{ov4} shows that if the Faddeev-Popov operator $\EuScript{M}^{ab}(A)\equiv-\partial_{\mu}D^{ab}_{\mu}(A)$ develops zero-modes, then the Landau gauge is not ideal. Gribov proved in \cite{Gribov:1977wm} that the operator $\EuScript{M}^{ab}(A)\equiv-\partial_{\mu}D^{ab}_{\mu}(A)$  does exhibit  in fact zero-modes. As a consequence, a residual gauge symmetry remains even after the implementation of \eqref{ov1}. The existence of such spurious configurations known as Gribov copies is the so-called \textit{Gribov problem}. For a pedagogical review of this subject, we refer to \cite{Sobreiro:2005ec,Vandersickel:2012tz,Vandersickel:2011zc,Pereira:2016inn}. Let us emphasize that the previous argument is restricted to Gribov copies generated by infinitesimal gauge transformations. Finite gauge transformations were considered in \cite{vanBaal:1991zw}.

The aforementioned discussion on the existence of the Gribov problem might induce the reader to think that this is a particular pathology of the Landau gauge or of a subclass of gauges, which can be circumvented by a suitable choice of a more appropriate gauge condition. Nevertheless, it was proved by Singer in \cite{Singer:1978dk} that this is not the case. In fact, it turns out that the Gribov problem has to do with the non-trivial topological structure of Yang-Mills theories.

In order to deal with the Gribov copies in the path integral measure, Gribov proposed in \cite{Gribov:1977wm} the restriction of the path integral domain to a smaller region $\Omega$ in field space, known as the Gribov region, defined as

\begin{equation}
\Omega = \left\{A^{a}_{\mu}\,,\,\partial_{\mu}A^{a}_{\mu}=0\,\Big|\,\EuScript{M}^{ab}(A)>0\right\}\,,
\label{ov5}
\end{equation}
namely, $\Omega$ is the set of gauge fields which satisfy the Landau gauge condition and for which the  Faddeev-Popov operator is strictly positive. The boundary $\partial\Omega$, where the first vanishing eigenvalue of the  Faddeev-Popov operator shows up, is called the first Gribov \textit{horizon}. We should mention that, although the region $\Omega$ is free from infinitesimal Gribov copies, it still contains additional copies \cite{vanBaal:1991zw} related to finite gauge transformations. A smaller region, contained inside $\Omega$ and known as the Fundamental Modular Region, exists and turns out to be fully free from Gribov ambiguities. However, unlike the Gribov region $\Omega$, a practical way to implement the restriction of the domain of integration in the functional integral to the Fundamental Modular Region has not yet been achieved so far. Therefore, we stick to the Gribov region $\Omega$ which displays important  properties: \textit{i)} it is bounded in all directions in field space; \textit{ii)} it is convex; \textit{iii)} all gauge orbits cross it at least once. These properties were proven in a rigorous fashion in \cite{Dell'Antonio:1991xt} and give a well defined support to original Gribov's proposal of restricting the functional integral to $\Omega$. Such restriction is effectively implemented through the addition of an extra term into the action \eqref{ov1},  as shown independently by Gribov and Zwanziger, \cite{Gribov:1977wm,Zwanziger:1989mf,Capri:2012wx}, {\it i.e.} 

\begin{equation}
\EuScript{Z}=\int_{\Omega} \left[\EuScript{D}A\right]\delta(\partial_{\mu}A^{a}_{\mu})|\mathrm{det}(\EuScript{M}^{ab})|\mathrm{e}^{-S_{\mathrm{YM}}}=\int  \left[\EuScript{D}A\right] \left[\EuScript{D}c\right] \left[\EuScript{D}\bar{c}\right]\left[\EuScript{D}b\right]\mathrm{e}^{-S_{\mathrm{FP}}-\gamma^4H(A)+dV\gamma^4(N^2-1)}\,,
\label{ov6}
\end{equation} 
where $H(A)$ is known as the \textit{horizon} function, being given by  

\begin{equation}
H(A)=g^2\int \dd^dx\dd^dy~f^{abc}A^{b}_{\mu}(x)\left[\EuScript{M}^{-1}(A)\right]^{ad}(x,y)f^{dec}A^{e}_{\mu}(y)\,.
\label{ov7}
\end{equation} 
In expression \eqref{ov6}, $V$ is the space-time volume and $\gamma$ is a parameter with the dimension of a mass, known as the \textit{Gribov parameter}. It is not a free parameter, being  determined in a self-consistent way through the so-called horizon condition, {\it i.e.}

\begin{equation}
\langle H(A)\rangle=dV(N^2-1)\,,
\label{ov8}
\end{equation}
where the expectation value $\langle\ldots\rangle$ is taken with respect to the modified measure given by eq.\eqref{ov6}.

As is apparent  from eq.\eqref{ov7}, the presence of the inverse of the Faddeev-Popov operator makes  the horizon function a  non-local quantity. Nevertheless, it can be cast in local form by the introduction of a suitable set of auxiliary fields, namely a pair of commuting $(\varphi^{ab}_{\mu},\bar{\varphi}^{ab}_{\mu})$ and of anticommuting $(\omega^{ab}_{\mu},\bar{\omega}^{ab}_{\mu})$ fields. Written in terms of these new fields, the Gribov-Zwanziger action $S_{\mathrm{GZ}}$ is expressed as

\begin{equation}
S_{\mathrm{GZ}}=S_{\mathrm{FP}}-\int \dd^dx\left(\bar{\varphi}^{ac}_{\mu}\EuScript{M}^{ab}{\varphi}^{bc}_{\mu}-\bar{\omega}^{ac}_\mu\EuScript{M}^{ab}\omega^{bc}_\mu+gf^{adl}\bar{\omega}^{ac}_\mu\partial_\nu\left(\varphi^{lc}_\mu D^{de}_\nu c^e\right)\right)+\gamma^{2}\int \dd^dx~gf^{abc}A^{a}_{\mu}(\varphi+\bar{\varphi})^{bc}_\mu\,,
\label{ov9}
\end{equation} 
and eq.\eqref{ov6} takes the form 

\begin{equation}
\EuScript{Z}=\int \left[\EuScript{D}\mu_{\mathrm{GZ}}\right]\mathrm{e}^{-S_{\mathrm{GZ}}+dV\gamma^4(N^2-1)}\,,
\label{ov10}
\end{equation}
with

\begin{equation}
\left[\EuScript{D}\mu_{\mathrm{GZ}}\right]=\left[\EuScript{D}A\right] \left[\EuScript{D}c\right] \left[\EuScript{D}\bar{c}\right]\left[\EuScript{D}b\right]\left[\EuScript{D}\varphi\right]\left[\EuScript{D}\bar{\varphi}\right]\left[\EuScript{D}\omega\right]\left[\EuScript{D}\bar{\omega}\right]\,.
\label{ov11}
\end{equation}

Remarkably, the action $S_{\mathrm{GZ}}$ is local and renormalizable to all orders in perturbation theory \cite{Zwanziger:1989mf}. Hence, the Gribov-Zwanziger action is an effective framework which implements the restriction of the domain of integration in the path integral  to the Gribov region $\Omega$ in a renormalizable and local way. 

The Gribov-Zwanziger action has many interesting and non-trivial properties. For our present purposes, we focus on a few of them. First, the gluon propagator computed out of \eqref{ov9} is suppressed in the deep infrared regime and attains a vanishing value at zero-momentum, a result which is at odds  with the divergent perturbative behavior. This propagator is said to be of the scaling type. Also, it violates reflection positivity and, therefore, gluons cannot be interpreted as excitations of  the physical spectrum, being thus confined. The ghost propagator, however, is enhanced in the strong coupling regime, diverging as $1/k^4$ for $k\approx 0$. Another  property  of the Gribov-Zwanziger action \eqref{ov9} is that it breaks  the BRST symmetry, given by the following transformations,

\begin{align}
sA^{a}_{\mu}&=-D^{ab}_{\mu}c^b\,,     &&sc^a=\frac{g}{2}f^{abc}c^bc^c\,, \nonumber\\
s\bar{c}^a&=b^{a}\,,     &&sb^{a}=0\,, \nonumber\\
s\varphi^{ab}_{\mu}&=\omega^{ab}_{\mu}\,,   &&s\omega^{ab}_{\mu}=0\,, \nonumber\\
s\bar{\omega}^{ab}_{\mu}&=\bar{\varphi}^{ab}_{\mu}\,,         &&s\bar{\varphi}^{ab}_{\mu}=0\,. 
\label{ov12}
\end{align}
in an explicit way, namely 

\begin{equation}
sS_{\mathrm{GZ}}=\gamma^{2}\int \dd^dx~gf^{abc}\left(-D^{ad}_{\mu}c^{d}(\varphi+\bar{\varphi})^{bc}_\mu+A^{a}_{\mu}\omega^{bc}_{\mu}\right)\,.
\label{ov13}
\end{equation}
Being proportional to $\gamma^2$, the BRST breaking is a soft breaking. It becomes relevant in the non-perturbative infrared region. Though, it does not  affect the deep ultraviolet region, so that  the perturbative results are recovered.  

More recently, it has been realized  that the localizing fields $(\varphi,\bar{\varphi},\omega,\bar{\omega})$ develop their own dynamics and non-trivial additional effects are generated. In particular, it has been  shown that dimension-two condensates, $\langle A^{a}_{\mu}A^{a}_{\mu}\rangle$ and $\langle\bar{\varphi}^{ab}_{\mu}\varphi^{ab}_{\mu}-\bar{\omega}^{ab}_{\mu}\omega^{ab}_{\mu}\rangle$, are dynamically generated \cite{Dudal:2007cw,Dudal:2008sp,Dudal:2011gd,Gracey:2010cg}, {\it i.e.}

\begin{equation}
\langle A^{a}_{\mu}A^{a}_{\mu}\rangle \propto \gamma^2\;, \,\,\,\, \,\,\,\,\,\,\langle\bar{\varphi}^{ab}_{\mu}\varphi^{ab}_{\mu}-\bar{\omega}^{ab}_{\mu}\omega^{ab}_{\mu}\rangle\propto \gamma^2\,.
\label{ov14}
\end{equation}
Taking into account the existence of such dimension-two condensates from the beginning, gives rise to the so-called Refined Gribov-Zwanziger action, which is expressed as

\begin{equation}
S_{\mathrm{RGZ}}=S_{\mathrm{GZ}}+\frac{m^2}{2}\int\dd^dx~A^{a}_{\mu}A^{a}_{\mu}-M^2\int \dd^dx\left(\bar{\varphi}^{ab}_{\mu}\varphi^{ab}_{\mu}-\bar{\omega}^{ab}_{\mu}\omega^{ab}_{\mu}\right)\,,
\label{ov15}
\end{equation}
where, much alike the Gribov parameter $\gamma^2$, the massive parameters $(m^2,M^2)$ are not independent, being determined by suitable gap equations obtained through the evaluation of the effective potential for the condensates $\langle A^{a}_{\mu}A^{a}_{\mu}\rangle$ and $\langle\bar{\varphi}^{ab}_{\mu}\varphi^{ab}_{\mu}-\bar{\omega}^{ab}_{\mu}\omega^{ab}_{\mu}\rangle$, see \cite{Dudal:2011gd}. 

The addition of the dimension-two operators, $A^{a}_{\mu}A^{a}_{\mu}$ and $(\bar{\varphi}^{ab}_{\mu}\varphi^{ab}_{\mu}-\bar{\omega}^{ab}_{\mu}\omega^{ab}_{\mu})$,  does not spoil the renormalizability of the refined action \eqref{ov15}. Notably, taking into account  these additional non-perturbative effects, changes the gluon and ghost propagators. For instance, the gluon propagator displays now  a decoupling/massive behavior, exhibiting a finite non-vanishing value at zero-momentum, while being still suppressed in the deep infrared sector. The ghost propagator, however, is not enhanced anymore in the strong coupling and,  for $k\approx 0$, it behaves as $1/k^2$. Such behavior of the gluon and ghost propagator is in very good agreement with the most recent lattice simulations in the Landau gauge, see \cite{Cucchieri:2007rg,Cucchieri:2008fc,Maas:2008ri,Cucchieri:2011ig,Oliveira:2012eh,Duarte:2016iko}.

An interesting property of the refinement of the Gribov-Zwanziger action is that its occurrence depends on the space-time dimension $d$. In particular, for $d=3,4$, the formation of dimension-two condensates is dynamically favoured  and the Gribov-Zwanziger action is naturally refined \cite{Dudal:2008sp,Dudal:2008rm}. Nevertheless, in $d=2$, infrared singularities prevent the introduction of such operators and the refinement does not take place. In particular, this implies that, for $d=3,4$, the gluon propagator is of decoupling type, while in $d=2$, it is of scaling type \cite{Dudal:2008xd}. Remarkably, this phenomenon was observed by recent lattice numerical simulationsn \cite{Cucchieri:2009zt,Dudal:2012hb}. It is worth mentioning that, considering the Gribov-Zwanziger action as an effective action with an energy scale ultraviolet cutoff, it is possible to show that, at the strong coupling, the refinement is also favored in $d>4$, \cite{Guimaraes:2016okb}. 

For completeness, we display the form of the tree-level gluon propagator in\footnote{Due to the different values of $d$, one should keep in mind the different meanings of the space-time indices and mass dimensions.} $d=3,4$ 

\begin{equation} 
\langle A^a_\mu(k) A^b_\nu(-k) \rangle_{d=3,4} = \delta^{ab} \frac{k^2 + M^2}{(k^2+m^2)(k^2+M^2) + 2g^2N \gamma^4} \left( \delta_{\mu\nu} - \frac{k_\mu k_\nu}{k^2} \right)\,, 
\label{ov16} 
\end{equation}
and in $d=2$,

\begin{equation}
\langle A^a_\mu(k) A^b_\nu(-k) \rangle_{d=2}  = \delta^{ab} \frac{k^2 }{k^4 + 2g^2N \gamma^4} \left( \delta_{\mu\nu} - \frac{k_\mu k_\nu}{k^2} \right)\,. 
\label{ov17} 
\end{equation}

%%%%%%%%%%%%%%%%%%%%%%%%%%%%%%%%%%%%%%%%%%%%%%%%%%%%%%%%%%%%%%
\subsection{Going beyond the Landau gauge and the non-perturbative BRST symmetry: linear covariant gauges} \label{beyondLandau}
%%%%%%%%%%%%%%%%%%%%%%%%%%%%%%%%%%%%%%%%%%%%%%%%%%%%%%%%%%%%%%

Although not peculiar to the Landau gauge, Gribov copies are very difficult to be handled when one chooses a different gauge condition. The main reason is that in the Landau gauge, the transversality of the gauge field ensures that the Faddeev-Popov operator $\EuScript{M}^{ab}$ is Hermitean. As such, this operator has a real spectrum which meaningfully allows for a definition of a Gribov region $\Omega$, where it is positive. Nevertheless, in general, the Faddeev-Popov operator is not Hermitean. The lack of such a property hinders a direct and clear definition of what would be the  Gribov region in gauges different from the Landau gauge. There are two notable examples of gauge choices which also possess a Hermitean Faddeev-Popov operator: the maximal Abelian and Coulomb gauges. For such gauges, an explicit construction of the Gribov-Zwanziger action and its refinement was performed, see \cite{Dudal:2006ib,Capri:2006cz,Capri:2008ak,Capri:2008vk,Capri:2010an,Gongyo:2013rua,Capri:2015pfa,Grotowski:1999ay,Zwanziger:2006sc,Burgio:2008jr,Reinhardt:2008pr,Golterman:2012dx,Guimaraes:2015bra,Burgio:2016nad}. Despite of this fact, these gauges have their own peculiarities and the development of the Refined Gribov-Zwanziger scenario for them is not at the same level as in the Landau gauge.  

A natural extension of the Landau gauge, which preserves Lorentz and color covariance, is given by the so-called linear covariant gauges, whose corresponding gauge condition is written as

\begin{equation}
\partial_{\mu}A^{a}_{\mu}=\alpha b^a\,,
\label{npbrst1}
\end{equation}
with $\alpha$ a non-negative gauge parameter and $b^a$ being, at this level, a  given function. Clearly, if one sets $\alpha=0$, the Landau gauge is recovered. Infinitesimal Gribov copies in these gauges are characterized by the zero-modes equation,

\begin{equation}
\EuScript{M}^{ab}_{\mathrm{LCG}}(A)\xi^b=-\delta^{ab}\partial^2\xi^b+gf^{abc}A^{c}_{\mu}\partial_{\mu}\xi^b+gf^{abc}(\partial_{\mu}A^{c}_{\mu})\xi^b = 0\,,
\label{npbrst2}
\end{equation}
where, in contrast to Landau gauge, $\partial_{\mu}A^{a}_{\mu}\neq0$ in general. It is precisely the fact that the gauge field is not purely transverse in these gauges that spoils the Hermiticity of $\EuScript{M}^{ab}_{\mathrm{LCG}}$.

The lack of Hermiticity makes the definition of the analogue of the Gribov region in linear covariant gauges very difficult. The first strategy to circumvent this technical difficulty was to take $\alpha$ as an infinitesimal parameter \cite{Sobreiro:2005vn}, namely, the linear covariant gauge is taken as a small perturbation of the Landau gauge. As a consequence, it was proven in \cite{Sobreiro:2005vn} that, in this situation, one can restrict the transverse component of the gauge field $A^{T,a}_{\mu}=(\delta_{\mu\nu}-\partial_{\mu}\partial_{\nu}/\partial^2)A^{a}_{\nu}$ to the Gribov region $\Omega$ in the domain of integration in the path integral and \textit{all} infinitesimal Gribov copies are removed. In \cite{Capri:2015pja}, it was pointed out that the same strategy works for finite values of $\alpha$ with the exception of pathological infinitesimal Gribov copies, corresponding to  zero modes which are  not smooth functions of the gauge parameter $\alpha$. Hence, modulo a certain subclass of pathological copies, the restriction of the domain of integration  in the path integral to the region where the transverse component belongs to $\Omega$ removes the infinitesimal Gribov copies. The resulting action was expressed in local form \cite{Capri:2015pja} and its renormalizability proof to all orders in perturbation theory was achieved  in \cite{Capri:2016aif}. We also refer to \cite{Moshin:2015gsa}.   

Nevertheless, the presence of the gauge parameter $\alpha$ allows for an explicit check of the gauge independence of correlation functions of gauge invariant operators. In standard perturbation theory, this is controlled by the BRST symmetry. However, the soft breaking of the BRST symmetry in the (refined) Gribov-Zwanziger setup gives rise to non-trivial complications for such a task. Nevertheless, recently,  in \cite{Capri:2015ixa}, a reformulation of the Gribov-Zwanziger action in the Landau gauge in terms of a transverse and gauge invariant field\footnote{We refer to Appendix~A of \cite{Capri:2015ixa} for the construction of the gauge invariant field $A^{h,a}_{\mu}$.}, see \cite{Zwanziger:1990tn,Lavelle:1995ty,Lavelle:2011yc}, $A^{h,a}_{\mu}$,  with $\partial_{\mu}A^{h,a}_{\mu}=0$, and its generalization to linear covariant gauges was proposed. In this new formulation, the  Gribov-Zwanziger enjoys an exact nilpotent BRST symmetry, which is a direct consequence of the gauge invariance of $A^{h,a}_\mu$ and which enables us to establish the independence from the parameter $\alpha$ of the gauge invariant correlation functions, and this even in the presence of the Gribov horizon.  

As shown in Appendix~A of \cite{Capri:2015ixa}, the gauge invariant field\footnote{We write it in the matrix notation $A^{h}_{\mu}=A^{h,a}_{\mu}T^a$, with $T^a$ the generators of $SU(N)$.} $A^{h}_{\mu}$ is expressed as an infinite series in powers of $A_\mu$, namely 

\begin{equation}
A^{h}_{\mu}=\left(\delta_{\mu\nu}-\frac{\partial_{\mu}\partial_{\nu}}{\partial^2}\right)\left(A_{\nu}-ig\left[\frac{1}{\partial^2}\partial A,A_\nu\right]+\frac{ig}{2}\left[\frac{1}{\partial^2}\partial A,\partial_{\nu}\frac{1}{\partial^2}\partial A\right]+\mathcal{O}(A^3)\right)\,,
\label{npbrst3}
\end{equation}
which, albeit transverse and gauge invariant, is a non-local expression. Upon a suitable redefinition of the field $b^a$, $b^a \rightarrow b^{h,a}$  \cite{Capri:2015ixa},  with the introduction of the gauge invariant field $A^{h,a}_{\mu}$, the resulting Gribov-Zwanziger action in linear covariant gauges  is written as \cite{Capri:2015ixa}

\begin{equation}
\tilde{S}^{\mathrm{LCG}}_{\mathrm{GZ}}=S_{\mathrm{YM}}+\int\dd^dx\left(b^{h,a}\partial_{\mu}A^{a}_{\mu}-\frac{\alpha}{2}b^{h,a}b^{h,a}+\bar{c}^{a}\partial_{\mu}D^{ab}_{\mu}(A)c^{b}\right)+\gamma^4H(A^h)\,,
\label{npbrst4}
\end{equation}
with

\begin{equation}
H(A^h)=g^2\int \dd^dx\dd^dy~f^{abc}A^{h,b}_{\mu}(x)\left[\EuScript{M}^{-1}(A^h)\right]^{ad}(x,y)f^{dec}A^{h,e}_{\mu}(y)\,,
\label{npbrst5}
\end{equation}
and

\begin{equation}
\EuScript{M}^{ab}(A^h)=-\delta^{ab}\partial^2+gf^{abc}A^{h,c}_{\mu}\partial_{\mu}\,,\,\,\,\,\mathrm{with}\,\,\,\,\,\partial_{\mu}A^{h,a}_{\mu}=0\,.
\label{npbrst6}
\end{equation}
Before proceeding, one should note that the horizon function $H(A^h)$ has now two sources of non-localities: the first one is related to the inverse of the operator $\EuScript{M}(A^h)$, which is similar to the non-locality of the horizon function in the Landau gauge, see eq.\eqref{ov7}. The second source of non-locality is associated with the field $A^{h}_{\mu}$ itself, see eq.\eqref{npbrst3}. In order to localize the first type of non-locality present in \eqref{npbrst5}, one proceeds as in the Landau gauge and introduces the set of auxiliary fields $(\bar{\varphi},\varphi,\bar{\omega},\omega)^{ab}_{\mu}$, which gives rise to the following action\footnote{We omit the vacuum term $-dV\gamma^{4}(N^2-1)$.}

\begin{eqnarray}
S^{\mathrm{LCG}}_{\mathrm{GZ}}&=&  S_{\mathrm{YM}} + \int \dd^dx\left(b^{h,a}\left(\partial_{\mu}A^{a}_{\mu}-\frac{\alpha}{2}b^{h,a}\right)+\bar{c}^{a}\partial_{\mu}D^{ab}_{\mu}c^{b}\right)\nonumber\\
&+&\int \dd^dx\left(\bar{\varphi}^{ac}_{\mu}\left[\EuScript{M}(A^h)\right]^{ab}\varphi^{bc}_{\mu}-\bar{\omega}^{ac}_{\mu}\left[\EuScript{M}(A^h)\right]^{ab}\omega^{bc}_{\mu}+g\gamma^2f^{abc}A^{h,a}_{\mu}(\varphi+\bar{\varphi})^{bc}_{\mu}\right)\,.
\label{npbrst7}
\end{eqnarray}
Some properties of \eqref{npbrst7} are listed: \textit{i)} The action $S^{\mathrm{LCG}}_{\mathrm{GZ}}$ is non-local due to the presence of the field $A^h_{\mu}$; \textit{ii)} In the limit $\alpha\rightarrow 0$, {\it i.e.} $\partial_\mu A^a_\mu=0$, one has that $A^h_{\mu}\rightarrow A^{T}_{\mu}$ and the action \eqref{npbrst8} is equivalent to \eqref{ov9}, the Gribov-Zwanziger action in the Landau gauge; \textit{iii)} The action \eqref{npbrst7} enjoys an exact  nilpotent BRST symmetry defined by the following transformations,

\begin{align}
sA^{a}_{\mu}&=-D^{ab}_{\mu}c^b\,,     &&sc^a=\frac{g}{2}f^{abc}c^bc^c\,, \nonumber\\
s\bar{c}^a&=b^{h,a}\,,     &&sb^{h,a}=0\,, \nonumber\\
s\varphi^{ab}_{\mu}&=0\,,   &&s\omega^{ab}_{\mu}=0\,, \nonumber\\
s\bar{\omega}^{ab}_{\mu}&=0\,,         &&s\bar{\varphi}^{ab}_{\mu}=0\,,\nonumber \\
sA^{h,a}_\mu &=0\,, 
\label{npbrst8}
\end{align}
with 
\begin{equation} 
s S^{\mathrm{LCG}}_{\mathrm{GZ}} = 0 \;.  \label{exact_brst} 
\end{equation}

Up to now, we have presented a BRST invariant non-local action,  eq.\eqref{npbrst7}, which restricts the domain of integration in the path integral to a region free from a large set of Gribov copies. Moreover, as reported in \cite{Capri:2016aqq}, this action can be fully localized by means of the introduction of additional  auxiliary fields. In particular, the localization procedure worked out in \cite{Capri:2016aqq} relies on the introduction of an auxiliary Stueckelberg-type field $\xi^a$, namely 

\begin{equation}
h=\mathrm{e}^{ig\xi^a T^a}\equiv \mathrm{e}^{ig\xi}.
\label{npbrst11} 
\end{equation}
The field $A^h_{\mu}=A^{h,a}_{\mu}T^{a}$ is expressed in terms of the local field $\xi^a$ as 

\begin{equation}
A^{h}_{\mu}=h^{\dagger}A_{\mu}h+\frac{i}{g}h^{\dagger}\partial_{\mu}h\,.
\label{npbrst12}
\end{equation}

An important feature of $A^h$, as defined by \eqref{npbrst12}, is that it is gauge invariant, that is  
\begin{equation}
A^{h}_{\mu} \rightarrow A^{h}_{\mu} \;,
\end{equation}
as can be explicitly seen  through a  gauge transformation parametrized by the $SU(N)$ matrix $V$
\begin{equation}
A_\mu \rightarrow V^{\dagger} A_\mu V + \frac{i}{g} V^{\dagger} \partial_ \mu V
\;, \qquad h \rightarrow V^{\dagger} h     \;, \qquad h^{\dagger} \rightarrow h^{\dagger} V \;.
\end{equation}

Although non-polynomial, the field $A^{h}_\mu$ \eqref{npbrst12} is now a local field and can be expanded in terms of $\xi^a$, yielding
\begin{equation}
(A^{h})^{a}_{\mu}=A^{a}_{\mu}-D^{ab}_{\mu}\xi^{b}-\frac{g}{2}f^{abc}\xi^{b}D^{cd}_{\mu}\xi^{d}+\mathcal{O}(\xi^{3})\,.
\end{equation}

Also, we must impose that the  local field $A^{h}_{\mu}$, eq.\eqref{npbrst12},  is transverse, namely, $\partial_{\mu}A^{h}_{\mu}=0$. Solving the transversality condition for the local field $\xi^a$ field, we obtain back the non-local expression for $A^{h}_{\mu}$ of eq.\eqref{npbrst3}, see Appendix~A of \cite{Capri:2015ixa}. Therefore, besides introducing the field $\xi^a$, we should enforce the transversality of $A^h_\mu$ by means of a Lagrange multiplier $\tau^a$, a task  which can be accomplished by introducing in the action the term 

\begin{equation}
S_\tau = \int \dd^dx~\tau^a\partial_\mu (A^{h}_{\mu})^a\,. 
\label{npbrst13}
\end{equation} 
We are now ready to write down the local and non-perturbative BRST invariant Gribov-Zwanziger action in the linear covariant gauges,  \textit{i.e.}

\begin{eqnarray}
S^{\mathrm{loc}}_{\mathrm{GZ}}&=& S_{\mathrm{YM}}+\int\dd^dx\left(b^a\partial_{\mu}A^{a}_{\mu}-\frac{\alpha}{2}b^a b^a+\bar{c}^a\partial_{\mu}D^{ab}_{\mu}(A)c^b\right)+\int\dd^dx~\tau^a\partial_\mu (A^{h}_{\mu})^a\nonumber\\
&-&\int\dd^dx\left(\bar{\varphi}^{ac}_{\mu}\left[\EuScript{M}(A^h)\right]^{ab}\varphi^{bc}_{\mu}-\bar{\omega}^{ac}_{\mu}\left[\EuScript{M}(A^h)\right]^{ab}\omega^{bc}_{\mu}+g\gamma^2 f^{abc}(A^{h}_{\mu})^a(\bar{\varphi}+\varphi)^{bc}_{\mu}\right)\,.
\label{npbrst14}
\end{eqnarray}

The local action turns out to be renormalizable to all orders in perturbation theory \cite{Capri:2017}, while implementing the restriction of the domain of integration in the path integral to a region free from a large set of Gribov copies in the linear covariant gauges in a BRST-invariant way. Such a feature allows for a well-defined Slavnov-Taylor identity, through which the gauge parameter independence of gauge-invariant correlation functions can be established. An extensive analysis of these properties was carried out in \cite{Capri:2016aqq} and \cite{Capri:2016gut}. 

As discussed in Subsect.~\ref{gribovlandau}, the action \eqref{npbrst14} needs to be further refined, due to the dynamical formation of dimension two condensates. This fact was exploited in \cite{Capri:2015nzw} where it was verified that, as in the Landau gauge, the refinement of the Gribov-Zwanziger action occurs in $d=3,4$, while in $d=2$ it is forbidden due to the presence of infrared singularities which prevent the formation of the dimension two condensates. Hence, in $d=3,4$, the action \eqref{npbrst14} is replaced by its refined version

\begin{equation}
S^{\mathrm{loc}}_{\mathrm{GZ}} \,\, \longrightarrow \,\, S^{\mathrm{loc}}_{\mathrm{RGZ}} = S^{\mathrm{loc}}_{\mathrm{GZ}} + \frac{m^2}{2}\int \dd^dx~(A^{h}_\mu)^a(A^{h}_\mu)^a - M^2\int \dd^dx\left(\bar{\varphi}^{ab}_{\mu}\varphi^{ab}_{\mu}-\bar{\omega}^{ab}_{\mu}{\omega}^{ab}_{\mu}\right)\,.
\label{npbrst15}
\end{equation}
The tree-level gluon propagator computed out of \eqref{npbrst15} is given by

\begin{equation}
\langle A^{a}_{\mu}(k)A^{b}_{\nu}(-k)\rangle_{d=3,4}=\delta^{ab}\left[\frac{k^2+M^2}{(k^2+m^2)(k^2+M^2)+2g^2\gamma^4N}\left(\delta_{\mu\nu}-\frac{k_{\mu}k_{\nu}}{k^2}\right)+\frac{\alpha}{k^2}\frac{k_{\mu}k_{\nu}}{k^2}\right]\,,
\label{npbrst16}
\end{equation}
being  in very good agreement with the most recent lattice data \cite{Cucchieri:2009kk,Cucchieri:2011pp,Bicudo:2015rma}. Although the transverse part of the propagator might acquire loop corrections, the longitudinal sector is exact to all orders, a consequence of the BRST symmetry. It is worth mentioning that this propagator has a decoupling/massive behavior in $d=3,4$, while in $d=2$ it is of scaling type due to the absence of refinement, see \cite{Capri:2015nzw}. For completeness, the local Refined Gribov-Zwanziger action, eq.\eqref{npbrst15}, is invariant under the nilpotent BRST transformations

\begin{align}
sA^{a}_{\mu}&=-D^{ab}_{\mu}c^b\,,     &&sc^a=\frac{g}{2}f^{abc}c^bc^c\,, \nonumber\\
s\bar{c}^a&=b^{h,a}\,,     &&sb^{h,a}=0\,, \nonumber\\
s\varphi^{ab}_{\mu}&=0\,,   &&s\omega^{ab}_{\mu}=0\,, \nonumber\\
s\bar{\omega}^{ab}_{\mu}&=0\,,         &&s\bar{\varphi}^{ab}_{\mu}=0\,,\nonumber \\
s h^{ij}& = -ig c^a (T^a)^{ik} h^{kj}  \;, && sA^{h,a}_\mu =0\,,   \nonumber \\
s\tau^a& =0\,, && s^2=0 \,,
\label{npbrst16a}
\end{align}

%\begin{equation}
%s h^{ij} = -ig c^a (T^a)^{ik} h^{kj}  \;, \qquad s (A^h)^a_\mu = 0  \;,  \label{brstst}
%\end{equation}
%
from which the BRST transformation of the field $\xi^a$,  eq.\eqref{npbrst11}, can be evaluated iteratively, giving
\begin{equation}
s \xi^a=  - c^a + \frac{g}{2} f^{abc}c^b \xi^c - \frac{g^2}{12} f^{amr} f^{mpq} c^p \xi^q \xi^r + O(g^3)    \;.
\label{eqsxi}
\end{equation}
It is instructive to check here explicitly the BRST invariance of $A^h$. For this, it is better to employ a
matrix notation for the fields, namely
\begin{eqnarray}
sA_\mu &=& -\partial_\mu c + ig [A_\mu, c] \;, \qquad s c = -ig c c     \;, \nonumber \\
s h & =& -igch \;, \qquad sh^{\dagger}  = ig h^{\dagger} c  \;, \label{mbrst}
\end{eqnarray}
with $A_\mu = A^a_\mu T^a$, $c=c^a T^a$, $\xi=\xi^a T^a$. From expression  \eqref{npbrst12} we get
\begin{eqnarray}
s A^h_\mu & = & ig h^{\dagger} c \;A_\mu h + h^{\dagger}  (-\partial_\mu c + ig [A_\mu, c]) h -ig h^{\dagger} A_\mu \;c h - h^{\dagger} c \partial_\mu h + h^{\dagger} \partial_\mu(c h) \nonumber \\
&=& igh^{\dagger} c A_\mu h -  h^{\dagger}  (\partial_\mu c ) h +ig  h^{\dagger} A_\mu \;c h - ig  h^{\dagger} c \;A_\mu h -ig h^{\dagger} A_\mu c h - h^{\dagger} c \partial_\mu h + h^{\dagger} (\partial_\mu c) h + h^{\dagger} c \partial_\mu h   \nonumber \\
&=& 0 \;. \label{sah}
\end{eqnarray}
Finally, we have  
\begin{equation} 
s S^{\mathrm{loc}}_{\mathrm{RGZ}} = 0 \;. \label{exr} 
\end{equation}
It is important to emphasize that, in the action \eqref{npbrst15},  the massive parameters $(\gamma,m,M)$ are coupled to BRST invariant expressions which are easily verified to be not BRST exact, {\it i.e.} cannot  be expressed as pure $s$-variations.  This fact ensures that these parameters are not akin to gauge parameters, having a physical meaning. As such, they will  be present in the gauge-invariant correlation functions. Also, they are not free, being determined by their own gap equations as discussed in \cite{Dudal:2008sp,Dudal:2011gd}.

%%%%%%%%%%%%%%%%%%%%%%%%%%%%%%%%%%%%%%%%%%%%%%%%%%%%%%%%%%%%%%
\subsection{Curci-Ferrari gauge} \label{CFgauge}
%%%%%%%%%%%%%%%%%%%%%%%%%%%%%%%%%%%%%%%%%%%%%%%%%%%%%%%%%%%%%%

In \cite{Pereira:2016fpn}, it was argued that the Gribov problem in the Curci-Ferrari gauge is intimately related to the existence of copies in the linear covariant gauges. By a suitable shift of the $b$-field, it was shown that the copies equation is the same in both gauges. As such, the issue of the Gribov copies can be handled in  the same  way and the implementation of the restriction of the domain of integration in the path integral is obtained by the introduction of the same horizon function \eqref{npbrst5}. As discussed in \cite{Pereira:2016fpn}, the Gribov-Zwanziger action in the Curci-Ferrari gauge is

\begin{eqnarray}
S^{\mathrm{CF}}_{\mathrm{GZ}}&=& S_{\mathrm{YM}}+\int \dd^dx\left[b^{h,a}\partial_{\mu}A^{a}_{\mu}+\bar{c}^{a}\partial_{\mu}D^{ab}_{\mu}c^{b}-\frac{\alpha}{2}b^{h,a}b^{h,a}+\frac{\alpha}{2}gf^{abc}b^{h,a}\bar{c}^{b}c^{c}
+\frac{\alpha}{8}g^{2}f^{abc}f^{cde}\bar{c}^{a}\bar{c}^{b}c^{d}c^{e}\right] \nonumber\\
&+&\int \dd^dx\left(\bar{\varphi}^{ac}_{\mu}\left[\EuScript{M}(A^h)\right]^{ab}\varphi^{bc}_{\mu}-\bar{\omega}^{ac}_{\mu}\left[\EuScript{M}(A^h)\right]^{ab}\omega^{bc}_{\mu}+g\gamma^2f^{abc}A^{h,a}_{\mu}(\varphi+\bar{\varphi})^{bc}_{\mu}\right)+\int\dd^dx~\tau^a\partial_\mu (A^{h}_{\mu})^a\,.
\label{npbrst17}
\end{eqnarray} 
As in the case of the linear covariant gauges, this theory suffers from non-perturbative  instabilities which give rise to the dynamical formation of condensates in $d=3,4$. Therefore, expression \eqref{npbrst17}  is refined by the inclusion of the same operators as in eq.\eqref{npbrst15} \textit{i.e.}

\begin{equation}
S^{\mathrm{CF}}_{\mathrm{GZ}} \,\, \longrightarrow \,\, S^{\mathrm{CF}}_{\mathrm{RGZ}} = S^{\mathrm{CF}}_{\mathrm{GZ}} + \frac{m^2}{2}\int \dd^dx~(A^{h}_\mu)^a(A^{h}_\mu)^a - M^2\int \dd^dx\left(\bar{\varphi}^{ab}_{\mu}\varphi^{ab}_{\mu}-\bar{\omega}^{ab}_{\mu}{\omega}^{ab}_{\mu}\right)\,.
\label{npbrst18}
\end{equation}
This action is invariant under the BRST transformations of eq.\eqref{npbrst16a}. The resulting tree-level gluon propagator coincides with that given in expression \eqref{npbrst16}. Nevertheless, since the Curci-Ferrari gauge is non-linear\footnote{The non-linearity of the Curci-Ferrari gauge can be appreciated through the fact that, upon elimination of the Lagrange multiplier field $b^a$,  a quartic ghost interaction term shows up.}, it does not enjoy the same set of Ward identities as the linear covariant gauges. A particular consequence of this fact is that, unlike the case of the linear covariant gauge,  the longitudinal part of the propagator is now affected by quantum corrections. 

%%%%%%%%%%%%%%%%%%%%%%%%%%%%%%%%%%%%%%%%%%%%%%%%%%%%%%%%%%%%%%
\subsection{Maximal Abelian gauge (MAG)}
%%%%%%%%%%%%%%%%%%%%%%%%%%%%%%%%%%%%%%%%%%%%%%%%%%%%%%%%%%%%%%

In order to construct the BRST-invariant (Refined) Gribov-Zwanziger action in the MAG, let us first  set our conventions for this gauge. To avoid unnecessary complications, we restrict ourselves to the case of the gauge group $SU(2)$. In this case, the gauge field $A_\mu=A^a_\mu T^a$ can be decomposed into diagonal and off-diagonal components, as

\begin{equation}
A_\mu = A^{a}_\mu T^a = A^{\alpha}_\mu T^\alpha + A^3_{\mu}T^3\,,
\label{mag1}
\end{equation}
with $\alpha=\left\{1,2\right\}$ denoting the indices corresponding to the off-diagonal components. The  diagonal generator $T^3\equiv T$ belongs to the Cartan subalgebra of $SU(2)$. Therefore, the following commutation relations hold:

\begin{eqnarray}
\left[T^a,T^b\right]&=&i\epsilon^{abc}T^c\,,\nonumber\\
\left[T^\alpha,T^\beta\right]&=&i\epsilon^{\alpha\beta 3}T^3\equiv i\epsilon^{\alpha\beta}T\,,\nonumber\\
\left[T^\alpha,T\right]&=&-i\epsilon^{\alpha\beta}T^\beta\,,\nonumber\\
\left[T,T\right]&=&0\,,
\label{mag2}
\end{eqnarray}
with $\epsilon^{\alpha\beta}= \epsilon^{\alpha\beta 3}$ being the totally antisymmetric symbol. The explicit decomposition of the field strength $F^a_{\mu\nu}$ yields

\begin{eqnarray}
F^{\alpha}_{\mu\nu} &=& \mathcal{D}^{\alpha\beta}_\mu A^{\beta}_\nu-\mathcal{D}^{\alpha\beta}_\nu A^{\beta}_\mu\,,\nonumber\\
F_{\mu\nu} &=& \partial_\mu A_\nu - \partial_\nu A_\mu + g\epsilon^{\alpha\beta}A^{\alpha}_\mu A^{\beta}_\nu\,,
\label{mag3}
\end{eqnarray}
with $\mathcal{D}^{\alpha\beta}_\mu$ being the covariant derivative defined with respect to the Abelian component $A_\mu=A_\mu^3$, namely

\begin{equation}
\mathcal{D}^{\alpha\beta}_\mu=\delta^{\alpha\beta}\partial_\mu-g\epsilon^{\alpha\beta}A_{\mu}\,.
\label{mag4}
\end{equation}

By means of  eq.\eqref{mag3}, we can express the Yang-Mills action as

\begin{equation}
S_{\mathrm{YM}}=\frac{1}{4}\int \dd^dx\left(F^{\alpha}_{\mu\nu}F^{\alpha}_{\mu\nu}+F_{\mu\nu}F_{\mu\nu}\right)\,,
\label{mag5}
\end{equation}
which is left invariant under the following infinitesimal gauge transformations,

\begin{eqnarray}
\delta A^{\alpha}_\mu &=& -\mathcal{D}^{\alpha\beta}_\mu \xi^\beta - g\epsilon^{\alpha\beta}A^{\beta}_{\mu}\xi\,,\nonumber\\
\delta A_\mu &=& -\partial_\mu \xi - g\epsilon^{\alpha\beta}A^{\alpha}_\mu \xi^\beta\,.
\label{mag6}
\end{eqnarray}
The MAG is defined by the gauge conditions,

\begin{eqnarray}
\mathcal{D}^{\alpha\beta}_\mu A^{\beta}_{\mu} &=& 0\,,\nonumber\\
\partial_\mu A_\mu &=& 0\,,
\label{mag7}
\end{eqnarray}
giving rise to the following Faddeev-Popov operator $\mathcal{M}^{\alpha\beta}(A)$: 

\begin{equation}
\mathcal{M}^{\alpha\beta}(A) = - \mathcal{D}^{\alpha \delta}_\mu \mathcal{D}^{\delta \beta}_\mu - g^2\epsilon^{\alpha\delta}\epsilon^{\beta\sigma}A^{\delta}_\mu A^{\sigma}_{\mu}\,.
\label{mag8}
\end{equation}

The gauge fixed Yang-Mills action in the MAG is written as

\begin{equation}
S^{\mathrm{FP}}_{\mathrm{MAG}} = S_{\mathrm{YM}}+\int \dd^dx\left(b^\alpha \mathcal{D}^{\alpha\beta}_\mu A^\beta_\mu - \bar{c}^{\alpha}\mathcal{M}^{\alpha\beta}(A)c^\beta + g\epsilon^{\alpha\beta}\bar{c}^{\alpha}(\mathcal{D}^{\alpha\delta}_{\mu}A^{\delta}_{\mu})c+b\partial_\mu A_\mu +\bar{c}\partial_\mu (\partial_\mu c + g\epsilon^{\alpha\beta}A^\alpha_\mu c^\beta)\right)\,.
\label{mag9}
\end{equation}
As discussed in \cite{Capri:2015pfa}, an analogous of the Gribov region $\Omega$ of the Landau gauge can be introduced in the MAG. More precisely, the Gribov region $\Omega_{\mathrm{MAG}}$ for the MAG is defined by 

\begin{equation}
\Omega_{\mathrm{MAG}} = \left\{\,A^\alpha_\mu\,,\,A_\mu\,;\,\mathcal{D}^{\alpha\beta}_\mu A^{\beta}_\mu=0\,,\,\partial_\mu A_\mu=0\,\Big|\,\mathcal{M}^{\alpha\beta}(A^h)>0\,\right\}\,.
\label{mag9.1}
\end{equation}
As in the case of the Landau gauge, the restriction of the domain of integration in the path integral to the region $\Omega_{\mathrm{MAG}}$ can be achieved in a BRST-invariant way by the introduction of the following horizon function

\begin{equation}
H_{\mathrm{MAG}}(A^h)=g^2\int \dd^dx \dd^dy~ A^{h,3}_\mu (x) \epsilon^{\alpha\beta}\left[\mathcal{M}^{-1}(A^h)\right]^{\alpha\delta}(x,y)\epsilon^{\delta\beta}A^{h,3}_\mu (y)\,,
\label{mag10}
\end{equation}
where $\mathcal{M}^{\alpha\beta}(A^h)$ means

\begin{equation}
\mathcal{M}^{\alpha\beta}(A^h) = - \mathcal{D}^{\alpha \delta}_\mu (A^h) \mathcal{D}^{\delta \beta}_\mu (A^h) - g^2\epsilon^{\alpha\delta}\epsilon^{\beta\sigma}A^{h,\delta}_\mu A^{h,\sigma}_{\mu}\,,
\label{mag11}
\end{equation}
and $\mathcal{D}^{\alpha\beta}_\mu (A^h) = \delta^{\alpha\beta}\partial_\mu - g\epsilon^{\alpha\beta}A^{h,3}_\mu$. The Gribov-Zwanziger action in the MAG is thus given by

\begin{equation}
\tilde{S}^{\mathrm{MAG}}_{\mathrm{GZ}} = S^{\mathrm{FP}}_{\mathrm{MAG}} + \gamma^4 H_{\mathrm{MAG}}(A^h)\,.
\label{mag12}
\end{equation}
As before, expression \eqref{mag12} has two sort of non-localities encoded in the horizon function $H_{\mathrm{MAG}}(A^h)$. In complete analogy with the procedure described in Subsect.~\ref{beyondLandau}, it is possible to cast the action \eqref{mag12} in a local fashion. The resulting local action is expressed by

\begin{eqnarray}
S^{\mathrm{MAG}}_{\mathrm{GZ}} &=& S^{\mathrm{FP}}_{\mathrm{MAG}} - \int \dd^dx\left(\bar{\varphi}^{\alpha\delta}_{\mu}\mathcal{M}^{\alpha\beta}(A^h)\varphi^{\beta\delta}_\mu-\bar{\omega}^{\alpha\delta}_\mu\mathcal{M}^{\delta\beta}(A^h)\omega^{\delta\beta}_\mu-g\gamma^2\epsilon^{\alpha\beta}A^{h,3}_{\mu}(\varphi+\bar{\varphi})^{\alpha\beta}_{\mu}\right)\nonumber\\
&+& \int \dd^dx\left(\tau^\alpha\partial_\mu A^{h,\alpha}_\mu+\tau \partial_\mu A^{h,3}_\mu\right)\,.
\label{mag13}
\end{eqnarray}
This action is invariant under the following BRST tranformations,

\begin{align}
sA^{\alpha}_{\mu}&=-(\mathcal{D}^{\alpha\beta}_{\mu}c^\beta+g\epsilon^{\alpha\beta}A^{\beta}_{\mu}c)\,,     &&sA_\mu=-(\partial_\mu c+g\epsilon^{\alpha\beta}A^{\alpha}_{\mu}c^{\beta})\,, \nonumber\\
sc^\alpha &=g\epsilon^{\alpha\beta}c^{\beta}c\,,     &&sc=\frac{g}{2}\epsilon^{\alpha\beta}c^{\alpha}c^{\beta}\,, \nonumber\\
s\bar{c}^{\alpha}&=b^\alpha\,,   &&s\bar{c}=b\,, \nonumber\\
s\bar{\omega}^{\alpha\beta}_{\mu}&=0\,,         &&s\bar{\varphi}^{\alpha\beta}_{\mu}=0\,,\nonumber \\
s\omega^{\alpha\beta}_{\mu}&=0\,,           &&s\varphi^{\alpha\beta}_\mu=0\,,\nonumber \\  
s\tau^{\alpha} &=0\,,    &&s\tau=0\,,\nonumber \\
sA^{h,\alpha}_\mu &=0\,,    &&sA^{h,3}_\mu = 0\,,
\label{mag14}
\end{align}
with 

\begin{equation} 
s S^{\mathrm{MAG}}_{\mathrm{GZ}} = 0 \;. \label{exmag}
\end{equation}

As in the case of the gauges discussed before,  the Gribov-Zwanziger action in the MAG also suffers from non-perturbative instabilities and dimension two condensates are dynamically generated in $d=3,4$, while,  in $d=2$, their formation is invalidated by infrared singularities. Therefore, as in the case of the previous  gauges,  the Gribov-Zwanziger action in the MAG does not refine in $d=2$.  We refer to \cite{Capri:2015pfa} for a detailed discussion of this feature. The Refined Gribov-Zwanziger action in $d=3,4$ in the MAG is written as

\begin{equation}
S^{\mathrm{MAG}}_{\mathrm{GZ}}\,\longrightarrow\,S^{\mathrm{MAG}}_{\mathrm{RGZ}}= S^{\mathrm{MAG}}_{\mathrm{GZ}} + \frac{m^2_{\mathrm{diag}}}{2}\int \dd^dx~A^{h,3}_{\mu}A^{h,3}_{\mu}+ \frac{m^2_{\mathrm{off}}}{2}\int \dd^dx~A^{h,\alpha}_{\mu}A^{h,\alpha}_{\mu}-M^2\int \dd^dx\left(\bar{\varphi}^{\alpha\beta}_{\mu}\varphi^{\alpha\beta}_{\mu}-\bar{\omega}^{\alpha\beta}_{\mu}\omega^{\alpha\beta}_{\mu}\right)\,,
\label{mag15}
\end{equation}
where the mass parameters  $(m^2_{\mathrm{diag}}, m^2_{\mathrm{off}}, M^2)$ reflect the existence of the dimension-two condensates $\langle  A^{h,3}_{\mu}A^{h,3}_{\mu}  \rangle$,  $\langle  A^{h,\alpha}_{\mu}A^{h,\alpha}_{\mu}  \rangle$,  $\langle \bar{\varphi}^{\alpha\beta}_{\mu}\varphi^{\alpha\beta}_{\mu}-\bar{\omega}^{\alpha\beta}_{\mu}\omega^{\alpha\beta}_{\mu} \rangle$. 

The diagonal gluon propagator is given by

\begin{equation}
\langle A_\mu (k)A_\nu (-k)\rangle = \left(\delta_{\mu\nu}-\frac{k_\mu k_\nu}{k^2}\right)\frac{k^2+M^2}{k^4+(m^2_{\mathrm{diag}}+M^2)k^2+M^2 m^2_{\mathrm{diag}}+4g^2\gamma^4}\,,
\label{mag16}
\end{equation}
while the off-diagonal gluon propagator is 

\begin{equation}
\langle A^{\alpha}_\mu (k) A^{\beta}_\nu (-k)\rangle = \left(\delta_{\mu\nu}-\frac{k_\mu k_\nu}{k^2}\right)\frac{\delta^{\alpha\beta}}{k^2+m^{2}_{\mathrm{off}}}\,.
\label{mag17}
\end{equation}
From expression \eqref{mag17}, we see that the off-diagonal gluon propagator displays a Yukawa type behavior. Lattice simulations give support to this result, see \cite{Bornyakov:2003ee,Mendes:2008ux,Gongyo:2012jb,Gongyo:2013sha}. Moreover, this behavior is in agreement with the Abelian dominance scenario \cite{Ezawa:1982bf}, where off-diagonal gluons should acquire a dynamical mass, reponsible for their decoupling at low energy. On the other hand, the diagonal gluon propagator \eqref{mag16} is of the refined Gribov type. As such, it is infrared suppressed and attains a non-vanishing value for $k=0$, in agreement with the lattice studies \cite{Bornyakov:2003ee,Mendes:2008ux,Gongyo:2012jb,Gongyo:2013sha}. The diagonal gluon propagator also displays reflection positivity violation, a feature which is interpreted as a signal of confinement. Again, this result is in agreement with the Abelian dominance scenario.

%%%%%%%%%%%%%%%%%%%%%%%%%%%%%%%%%%%%%%%%%%%%%%%%%%%%%%%%%%%%%%
\section{Non-perturbative coupling of scalar fields in the adjoint representation} \label{noscalaradjoint}
%%%%%%%%%%%%%%%%%%%%%%%%%%%%%%%%%%%%%%%%%%%%%%%%%%%%%%%%%%%%%%

In this section, we generalize  the construction of \cite{Capri:2014bsa} to linear covariant, Curci-Ferrari and maximal Abelian gauges. To begin with, we consider scalar fields in the adjoint representation of\footnote{In the case of the MAG, we restrict ourselves to $SU(2)$ for simplicity.} $SU(N)$. The idea proposed in \cite{Capri:2014bsa} consists in the introduction of a term akin to the horizon function for the matter sector, which provides a non-perturbative coupling between matter fields and the gauge sector. Although for the gluon sector  the horizon function has a clear geometrical meaning, implementing the restriction of the domain of integration in the path integral to the Gribov region, the introduction of an analogous term in the matter sector does not yet exhibit the same well defined geometric support. Nevertheless, recently, it was observed that such a non-perturbative coupling between matter and gauge fields could be motivated through the dimensional reduction of higher-dimensional Yang-Mills theory, see \cite{Guimaraes:2016okb}. More precisely, upon reduction of a five dimensional Yang-Mills to the four dimensional theory \cite{Guimaraes:2016okb}, a non-perturbative coupling between the scalar field corresponding to the fifth component of the gauge connection and the four dimensional gauge field shows up, being precisely of the type introduced in \cite{Capri:2014bsa}. As we shall see, this prescription gives rise to non-perturbative matter fields propagators which turn out to be in good agreement with lattice data, whenever available.

%%%%%%%%%%%%%%%%%%%%%%%%%%%%%%%%%%%%%%%%%%%%%%%%%%%%%%%%%%%%%%
\subsection{Linear covariant and Curci-Ferrari gauges} \label{scalarLCGCF}
%%%%%%%%%%%%%%%%%%%%%%%%%%%%%%%%%%%%%%%%%%%%%%%%%%%%%%%%%%%%%%

Let us consider the standard action of scalar fields in the adjoint representation of $SU(N)$, minimally coupled with the gauge sector, {\it i.e.} 

\begin{equation}
S_{\mathrm{scalar}}=\int \dd^dx\left[\frac{1}{2}(D^{ab}_\mu \phi^{b})(D^{ac}_\mu \phi^c)+\frac{m^2_\phi}{2}\phi^a \phi^a + \frac{\lambda}{4!}(\phi^a\phi^a)^2\right]\,.
\label{scalarlcg1}
\end{equation}
Of course, expression \eqref{scalarlcg1} is left  invariant by BRST transformations \eqref{npbrst8}, with the scalar field $\phi^a$ transforming as 
\begin{equation}
s \phi = ig [\phi, c]  \;, \qquad   \phi = \phi^a T^a  \;,
\end{equation} 
where $\{ T^a \}$ stand for the generators of $SU(N)$ in the adjoint representation. 

Making use of the Stueckelberg field $\xi$,  eq.\eqref{npbrst11},  a BRST invariant scalar field is constructed as follows \cite{Capri:2016aqq}: 
\begin{equation}
 \phi^h = h^{\dagger} \phi h  \;, \qquad h=e^{ig\,\xi^{a}T^{a}} \;.
\end{equation}
To first order, we get 
\begin{equation} 
\phi^{h,a} = \phi^a + g f^{abc} \xi^b \phi^c + O(\xi^2) \;. \label{fophi}
\end{equation}
It is easy to verify that $\phi^h$ is left invariant by the BRST transformations, {\it i.e.}
\begin{equation}
 s \phi^h = 0\,.
\end{equation}
The prescription introduced in \cite{Capri:2014bsa} amounts to introduce the following non-local BRST invariant term to the scalar action \eqref{scalarlcg1},

\begin{equation}
\EuScript{H}(\phi^h)=g^2\int \dd^dx \dd^dy~f^{abc}\phi^{h,b}(x)\left[\EuScript{M}^{-1}(A^h)\right]^{ad}(x,y)f^{dec}\phi^{h,e}(y)\,, 
\label{scalarlcg2}
\end{equation}
where  $\left( \EuScript{M}(A^h) \right)^{ad}$ stands for the Faddeev-Popov operator of eq.\eqref{npbrst6}. 
It is almost immediate to realise that expression \eqref{scalarlcg2}  shares great similarity with the horizon function of the gluon sector, eq.\eqref{npbrst5}. In fact, as already mentioned, expression \eqref{scalarlcg2} can be obtained through the dimensional reduction of higher-dimensional Yang-Mills theory \cite{Guimaraes:2016okb}.

%The field $\phi^{h,a}$ plays the analogue role of $A^{h,a}_\mu$. In particular, it is constructed to be gauge invariant. As in the case of $A^{h,a}_\mu$, it is non-local and has the following structure,

%\begin{equation}
%\phi^{h,a} = \phi^a + gf^{abc}\phi^c\frac{1}{\partial^2}\partial_\mu A^{b}_{\mu} + \mathcal{O}(A^2)\,.
%\label{scalarlcg3}
%\end{equation}
%Its gauge invariance is of most importance for the BRST-invariance of the theory. 

The action of the scalar field  with the addition of the non-perturbative coupling \eqref{scalarlcg2} is given by 

\begin{equation}
S^{\phi} = \int \dd^dx\left[\frac{1}{2}(D^{ab}_\mu \phi^{b})(D^{ac}_\mu \phi^c)+\frac{m^2_\phi}{2}\phi^a \phi^a + \frac{\lambda}{4!}(\phi^a\phi^a)^2\right] + \sigma^4 \EuScript{H}(\phi^h)\,,
\label{scalarlcg4}
\end{equation}
where the massive parameter $\sigma$ plays the same role of the Gribov parameter $\gamma$.  Again, due to the presence of the operator  $\EuScript{M}^{-1}$ in expression \eqref{scalarlcg2}, the action \eqref{scalarlcg4} is non-local. Moreover, it turns out to be possible to cast the action $S^{\phi}$  in local form following the same procedure adopted in the previous sections for the localization of the Gribov-Zwanziger action. To that purpose, we introduce a set of auxiliary fields $(\bar{\eta},\eta, \bar{\theta},\theta)^{ab}$ akin to Zwanziger's localizing fields in such a way that

\begin{equation}
\sigma^4\EuScript{H}(\phi^{h})\,\longrightarrow\,-\int \dd^dx\left(\bar{\eta}^{ac}\EuScript{M}^{ab}(A^h)\eta^{bc}-\bar{\theta}^{ac}\EuScript{M}^{ab}(A^h)\theta^{bc}-g\sigma^2f^{abc}\phi^{h,c}(\bar{\eta}+\eta)^{ab}\right)\,.
\label{scalarlcg5}
\end{equation}
The fields $(\bar{\eta},\eta)$ are commuting while $(\bar{\theta},\theta)$ are anti-commuting. Integrating out these fields  in the functional integration gives back the non-local  expression \eqref{scalarlcg2}. 

%Finally, the localization of $A^h_\mu$ is achieved by the introduction of a Stueckelberg type field as discussed in Subsect.~\ref{beyondLandau}. Finally, we should localize the gauge-invariant scalar field $\phi^h$. As discussed in \cite{Capri:2016aqq}, it is possible to express $\phi^{h,a}$ in terms of the Stueckelberg field $\xi^a$ by\footnote{Inhere, we employ the matrix notation $\phi^h=\phi^{h,a}T^a$.}

%\begin{equation}
%\phi^h = h^{\dagger}\phi h\,,
%\label{scalarlcg6}
%\end{equation}
%with $h$ given by eq.\eqref{npbrst11}. Solving the transversality condition $\partial_\mu A^h_\mu = 0$ for $\xi$ and plugging the result into eq.\eqref{scalarlcg6}, we obtain the non-local expression \eqref{scalarlcg3}. It is simple to check that the field $\phi^h$ is BRST-invariant, namely

%\begin{equation}
%s\phi^h=0\,.
%\label{scalarlcg7}
%\end{equation}
Therefore, the local scalar field action non-perturbatively coupled to the gauge sector is expressed by

\begin{eqnarray}
\tilde{S}^{\phi}_{\mathrm{loc}} &=& \int \dd^dx\left[\frac{1}{2}(D^{ab}_\mu \phi^{b})(D^{ac}_\mu \phi^c)+\frac{m^2_\phi}{2}\phi^a \phi^a + \frac{\lambda}{4!}(\phi^a\phi^a)^2\right]\nonumber\\
&-&\int \dd^dx\left(\bar{\eta}^{ac}\EuScript{M}^{ab}(A^h)\eta^{bc}-\bar{\theta}^{ac}\EuScript{M}^{ab}(A^h)\theta^{bc}-g\sigma^2f^{abc}\phi^{h,c}(\bar{\eta}+\eta)^{ab}\right)\,.
\label{scalarlcg8}
\end{eqnarray}
One should keep in mind that in expression \eqref{scalarlcg8}, both $A^h_\mu$ and $\phi^h$ are expressed in terms of the Stueckelberg field $\xi^a$ and are thus local fields, albeit non-polynomial. 

As it happens in the gauge sector of the Gribov-Zwanziger action, the non-local mass term  \eqref{scalarlcg2} entails non-perturbative instabilities which give rise to the dimension two condensates, $\langle \phi^{h,a} \phi^{h,a} \rangle$ and $\langle \bar{\eta}^{ab}\eta^{ab}-\bar{\theta}^{ab}\theta^{ab} \rangle$, akin to those of the Refined Gribov-Zwanziger action. It is worth to proceed by evaluating those condensates at first order, a task which can be accomplished by  introducing the operators

\begin{equation}
J\int \dd^dx~\phi^{h,a}\phi^{h,a}\,\,\,\,\,\,\mathrm{and}\,\,\,\,\,\,-\tilde{J}\int \dd^dx\left(\bar{\eta}^{ab}\eta^{ab}-\bar{\theta}^{ab}\theta^{ab}\right)\,,
\label{scalarlcg9}
\end{equation}
in expression \eqref{scalarlcg8}, where $(J,\tilde{J})$ are constant sources. Thus, we define the action $\Sigma(J,\tilde{J})$ by

\begin{equation}
\Sigma(J,\tilde{J}) = \tilde{S}^{\phi}_{\mathrm{loc}} + J\int \dd^dx~\phi^{h,a}\phi^{h,a}-\tilde{J}\int \dd^dx\left(\bar{\eta}^{ab}\eta^{ab}-\bar{\theta}^{ab}\theta^{ab}\right)\,.
\label{scalarlcg10}
\end{equation}
To first order, the condensates $\langle \phi^{h,a} \phi^{h,a} \rangle$ and $\langle \bar{\eta}^{ab}\eta^{ab}-\bar{\theta}^{ab}\theta^{ab} \rangle$ can be obtained by taking the derivatives of the one-loop vacuum energy $\mathcal{E}^{(1)}$ with respect to the sources $(J,\tilde{J})$, and setting them to zero, where

\begin{equation}
\mathrm{e}^{-V\mathcal{E}^{(1)}}=\int \left[\EuScript{D}\mu\right]\mathrm{e}^{-\Sigma^{(2)}(J,\tilde{J})}\,.
\label{scalarlcg12}
\end{equation}
$\Sigma^{(2)}(J,\tilde{J})$ denotes  the quadratic part of \eqref{scalarlcg10}, while  the path integral measure is expressed as

\begin{equation}
\left[\EuScript{D}\mu\right]=\left[\EuScript{D}A\right]\left[\EuScript{D}b\right]\left[\EuScript{D}\bar{c}\right]\left[\EuScript{D}c\right]\left[\EuScript{D}\bar{\omega}\right]\left[\EuScript{D}\omega\right]\left[\EuScript{D}\bar{\varphi}\right]\left[\EuScript{D}\varphi\right]\left[\EuScript{D}\xi\right]\left[\EuScript{D}\tau\right]\left[\EuScript{D}\phi\right]\left[\EuScript{D}\bar{\eta}\right]\left[\EuScript{D}\eta\right]\left[\EuScript{D}\bar{\theta}\right]\left[\EuScript{D}\theta\right]\,.
\label{scalarlcg13}
\end{equation}
At one-loop order, the vacuum energy is easily evaluated, being given by

\begin{equation}
\mathcal{E}^{(1)} = \frac{(N^2-1)}{2}\int \frac{\dd^dp}{(2\pi)^d}\mathrm{ln}\left(p^2+m^2_\phi+2J+\frac{2Ng^2\sigma^4}{p^2+\tilde{J}}\right)\,,
\label{scalarlcg14}
\end{equation}
where dimensional regularization has been  employed. Therefore, at first order, for the condensates we get 

\begin{eqnarray}
\langle \phi^{h,a} \phi^{h,a} \rangle &=& \frac{\partial\mathcal{E}^{(1)}}{\partial J}\Big|_{J=\tilde{J}=0}=-(N^2-1)\int\frac{\dd^dk}{(2\pi)^d}\frac{m^2_\phi}{k^4+m^2_\phi k^2+2Ng^2\sigma^4}\nonumber\\
&-&2Ng^2\sigma^4(N^2-1)\int \frac{\dd^dk}{(2\pi)^d}\frac{1}{k^2}\frac{1}{k^4+m^2_\phi k^2+2Ng^2\sigma^4}\,,\nonumber\\
\langle \bar{\eta}^{ab}\eta^{ab}-\bar{\theta}^{ab}\theta^{ab}\rangle &=& -\frac{\partial \mathcal{E}^{(1)}}{\partial \tilde{J}}\Big|_{J=\tilde{J}=0}= (N^2-1)Ng^2\sigma^4\int \frac{\dd^dk}{(2\pi)^d}\frac{1}{k^2}\frac{1}{k^4+m^2_\phi k^2 + 2Ng^2\sigma^4}\,.
\label{scalarlcg15}
\end{eqnarray}
One sees that the one-loop result already shows non-vanishing expressions for the condensates $\langle \phi^{h,a} \phi^{h,a} \rangle$ and $\langle \bar{\eta}^{ab}\eta^{ab}-\bar{\theta}^{ab}\theta^{ab} \rangle$. Remarkably, the contributions coming from the introduction of the non-perturbative mass term  \eqref{scalarlcg2} to the standard scalar field action are ultraviolet convergent. Interesting to note, very much alike the refinement of the Gribov-Zwanziger action, infrared singularities show up in the integrals \eqref{scalarlcg15}, preventing the formation of such condensates in $d=2$. As in the case of the gluon sector, in $d=3,4$, the effects of the existence of the  condensates   $\langle \phi^{h,a} \phi^{h,a} \rangle$ and $\langle \bar{\eta}^{ab}\eta^{ab}-\bar{\theta}^{ab}\theta^{ab} \rangle$ can be taken into account by refining the matter action as:

\begin{equation}
\tilde{S}^{\phi}_{\mathrm{loc}}\,\longrightarrow\, S^{\phi}_{\mathrm{loc}} = \tilde{S}^{\phi}_{\mathrm{loc}} + \tilde{m}^2_\phi\int \dd^dx~\phi^{h,a}\phi^{h,a}-\rho^2\int \dd^dx\left(\bar{\eta}^{ab}\eta^{ab}-\bar{\theta}^{ab}\theta^{ab}\right)\,,
\label{scalarlcg16}
\end{equation}
where the parameters $(\tilde{m}^2_\phi, \rho^2)$ have  dynamical origin and can be  obtained through the evaluation of the effective potential for $\langle \phi^{h,a} \phi^{h,a} \rangle$ and $\langle \bar{\eta}^{ab}\eta^{ab}-\bar{\theta}^{ab}\theta^{ab} \rangle$.  Furthermore, in the case of $d=2$, due to the absence of condensates, the action remains the original one given by eq.\eqref{scalarlcg8}. 

After these considerations, we can compute the tree-level scalar field propagator for different values of $d$. In $d=2$, we have
\begin{equation}
\langle \phi(k)\phi(-k)\rangle_{d=2} = \delta^{ab}\frac{k^2}{k^4+m^2_\phi k^2+2Ng^2\sigma^4}\,,
\label{scalarlcg17}
\end{equation}
while in $d=3,4$,

\begin{equation}
\langle \phi(k)\phi(-k)\rangle_{d=3,4} = \delta^{ab}\frac{k^2+\rho^2}{k^4+(m^2_\phi+\tilde{m}^2_\phi+\rho^2) k^2+(m^2_\phi+\tilde{m}^2_\phi)\rho^2+2Ng^2\sigma^4}\,. 
\label{scalarlcg18}
\end{equation}
In analogy with the case of gluon propagator, the scalar field propagator attains a finite value at zero momentum in $d=3,4$ while in $d=2$ it vanishes at $k=0$. In both cases the scalar propagator is infrared suppressed. Also, at the tree-level, there is no $\alpha$-dependence as it is apparent from eqs.\eqref{scalarlcg17} and \eqref{scalarlcg18}. Hence, the Landau limit $\alpha=0$ is trivial and agrees with the results reported in \cite{Dudal:2008sp}. Also, the propagators \eqref{scalarlcg17} and \eqref{scalarlcg18} violate reflection positivity, a feature which is interpreted as a signal of confinement. We see thus that the introduction of the  non-perturbative matter coupling \eqref{scalarlcg2} has the effect of confinining the scalar matter fields.

It is worth here to add some further remarks on the specific case of $d=2$, eq.\eqref{scalarlcg17}. One should keep in mind that expression \eqref{scalarlcg17}  is a consequence of the first-order absence of the condensate $\langle \bar{\eta}^{ab}\eta^{ab}-\bar{\theta}^{ab}\theta^{ab} \rangle$, as it follows from  eq.\eqref{scalarlcg15}. Though, we underline that this is only a first order analysis. As such, expression \eqref{scalarlcg17} would retain its validity at this order. Willing to make an all order statement, a higher loop analysis of the condensate 
$\langle \bar{\eta}^{ab}\eta^{ab}-\bar{\theta}^{ab}\theta^{ab} \rangle$ would be required, a matter which is well beyond the aim of the present paper. 
Although the available lattice simulations  \cite{Maas:2011yx} point towards a similar behavior for the scalar field propagator in the infrared for different values of $d=4,3,2$ in the Landau gauge, in order to make a comparison with the lattice data in $d=2$ a detailed analysis of the higher order condensate $\langle \bar{\eta}^{ab}\eta^{ab}-\bar{\theta}^{ab}\theta^{ab} \rangle$  would definitively be needed. 

Finally, as dicussed in Subsect.~\ref{CFgauge}, the Gribov problem in the Curci-Ferrari and linear covariant gauges can be treated by means of a formal equivalence. As such, the non-perturbative matter term, eq.\eqref{scalarlcg2},  in both gauges is the same. Therefore, the scalar field action non-perturbatively coupled to  the gauge sector is given by \eqref{scalarlcg8}. Clearly, at first order, all the computations presented in this section remain valid for the Curci-Ferrari gauges, namely, the calculation of the vacuum energy and of the scalar field propagator. Of course,  taking into account higher loops contributions, the non-linear character of the Curci-Ferrari gauges will show up giving results which will differ from those of the linear covariant gauges. Since this is beyond the scope of the present work, we limit ourselves to the first order computations already presented in the case of linear covariant gauges, which retain their validity also in the Curci-Ferrari gauges.

%%%%%%%%%%%%%%%%%%%%%%%%%%%%%%%%%%%%%%%%%%%%%%%%%%%%%%%%%%%%%%
\subsection{Maximal Abelian gauge}
%%%%%%%%%%%%%%%%%%%%%%%%%%%%%%%%%%%%%%%%%%%%%%%%%%%%%%%%%%%%%%

In the case of the MAG, although the prescription is the same, care is due to the decomposition of color indices into diagonal and off-diagonal ones. Firstly, we express the minimally coupled scalar field action in a color decomposed fashion, namely

\begin{eqnarray}
S_{\mathrm{scalar}} &=& \int \dd^dx~\frac{1}{2}\left\{(\partial_{\mu}\phi^\alpha)(\partial_{\mu}\phi^\alpha)+(\partial_\mu \phi)(\partial_\mu \phi) - 2g\epsilon^{\alpha\beta}\left[(\partial_\mu \phi)\phi^\alpha A^\beta_\mu-(\partial_\mu\phi^\alpha)\phi A^\beta_\mu+(\partial_\mu\phi^\alpha)\phi^\beta A_\mu\right]\right.\nonumber\\
&+&\left.g^2\left[A^{\alpha}_{\mu}A^{\alpha}_{\mu}(\phi^\beta\phi^\beta+\phi\phi)+A_\mu A_\mu\phi^\alpha\phi^\alpha - A^{\alpha}_\mu A^{\beta}_\mu\phi^a\phi^b-2A_\mu \phi A^\alpha_\mu \phi^\alpha\right]\right\}+\int \dd^dx\frac{m^2_\phi}{2}(\phi^\alpha\phi^\alpha+\phi \phi)\nonumber\\
&+& \int \dd^dx\frac{\lambda}{4!}\left[(\phi^\alpha\phi^\alpha)^2+2\phi^\alpha\phi^\alpha\phi^2+\phi\phi\phi\phi\right]\,,
\label{scalarmag1}
\end{eqnarray}
with $\phi\equiv\phi^3$. 

%For convenience, we write

%\begin{equation}
%\frac{m^2_\phi}{2}(\phi^\alpha \phi^\alpha + \phi\phi) \equiv \frac{\tilde{m}^2_{\mathrm{off}}}{2}\phi^\alpha\phi^\alpha+\frac{\tilde{m}^2_\mathrm{diag}}{2}\phi\phi\,.
%\label{scalarmag2}
%\end{equation}

As in the case of linear covariant and Curci-Ferrari gauges, the non-perturbative matter coupling is obtained through the addition, in the scalar field action, of a non-local term which shares great similarity with the corresponding horizon function of the gluon sector in the MAG, eq.\eqref{mag10}, {\it i.e.}

\begin{equation}
\mathcal{H}(\phi^h) = g^2\int \dd^dx \dd^dy~\epsilon^{\alpha\beta}\phi^{h,3}(x)\left[\mathcal{M}^{-1}(A^h)\right]^{\alpha\delta}(x,y)\epsilon^{\delta\beta}\phi^{h,3}(y)\,,
\label{scalarmag3}
\end{equation}
where the Faddeev-Popov operator  $\mathcal{M}(A^h)$ is now given by eq.\eqref{mag11}. The scalar field action supplemented with the non-perturbative coupling \eqref{scalarmag3} becomes

\begin{equation}
S^\phi_{\mathrm{MAG}} = S_\mathrm{scalar}+\sigma^4\mathcal{H}(\phi^h)\,.
\label{scalarmag4}
\end{equation} 
The parameter $\sigma$ has mass dimension and is the analogue of the Gribov parameter $\gamma$ in the matter sector. As before, the non-local action \eqref{scalarmag4} can be cast in local form by means of the introduction of auxiliary fields and of a Stueckelberg field, also used to localize $A^h_\mu$. In local form, the action \eqref{scalarmag4} is written as

\begin{equation}
\tilde{S}^\phi_{\mathrm{MAG-loc}} = S_\mathrm{scalar}-\int \dd^dx\left(\bar{\eta}^{\alpha\delta}\mathcal{M}^{\alpha\beta}(A^h)\eta^{\beta\delta}-\bar{\theta}^{\alpha\delta}\mathcal{M}^{\alpha\beta}(A^h)\theta^{\beta\delta}-g\sigma^2\epsilon^{\alpha\beta}\phi^{h,3}(\bar{\eta}+\eta)^{\alpha\beta}\right)\,,
\label{scalarmag5}
\end{equation}
with $\phi^h=h^\dagger \phi h$.

As pointed out in Subsect.~\ref{scalarLCGCF}, the auxiliary localizing fields $(\bar{\eta},\eta,\bar{\theta},\theta)^{\alpha\beta}$ develop their own dynamics and give rise to the dynamical formation of condensates. This is in very much analogy with the refinement of the Gribov-Zwanziger action. In order to explicitly check the existence of such condensates to first order, we proceed as before and introduce  the following operators to \eqref{scalarmag5},

\begin{equation}
J\int \dd^dx~\phi^{h,3}\phi^{h,3}\,\,\,\,\,\,\mathrm{and}\,\,\,\,\,\,-\tilde{J}\int \dd^dx\left(\bar{\eta}^{\alpha\beta}\eta^{\alpha\beta}-\bar{\theta}^{\alpha\beta}\theta^{\alpha\beta}\right)\,,
\label{scalarmag6}
\end{equation} 
where $J$ and $\tilde{J}$ are constant sources. This gives rise to

\begin{equation}
\Sigma (J,\tilde{J}) = \tilde{S}^\phi_{\mathrm{MAG-loc}}+J\int \dd^dx~\phi^{h,3}\phi^{h,3}-\tilde{J}\int \dd^dx\left(\bar{\eta}^{\alpha\beta}\eta^{\alpha\beta}-\bar{\theta}^{\alpha\beta}\theta^{\alpha\beta}\right)\,.
\label{scalarmag7}
\end{equation}
Our aim is to compute the following condensates at one-loop order:

\begin{equation}
\langle \phi^{h,3}(x) \phi^{h,3}(x)\rangle \,\,\,\,\,\,\mathrm{and}\,\,\,\,\,\, \langle \bar{\eta}^{\alpha\beta}(x)\eta^{\alpha\beta}(x) - \bar{\theta}^{\alpha\beta}(x)\theta^{\alpha\beta}(x)\rangle\,.
\label{scalarmag8}
\end{equation}
This is achieved by taking the derivatives with respect to $J$ and $\tilde{J}$ of the vacuum energy $\mathcal{E}$, defined by

\begin{equation}
\mathrm{e}^{-V\mathcal{E}(J,\tilde{J})} = \int \left[\EuScript{D}\mu\right]\mathrm{e}^{-S^{\mathrm{loc}}_{\mathrm{RGZ}}-\Sigma (J,\tilde{J})}\,,
\label{scalarmag9}
\end{equation}
and setting the sources to zero at the end, namely 

\begin{eqnarray}
\langle \phi^{h,3}(x) \phi^{h,3}(x)\rangle &=& \frac{\partial\mathcal{E}(J,\tilde{J})}{\partial J}\Big|_{J=\tilde{J}=0}\,,\nonumber\\
\langle \bar{\eta}^{\alpha\beta}(x)\eta^{\alpha\beta}(x) - \bar{\theta}^{\alpha\beta}(x)\theta^{\alpha\beta}(x)\rangle &=& - \frac{\partial\mathcal{E}(J,\tilde{J})}{\partial \tilde{J}}\Big|_{J=\tilde{J}=0}\,.
\label{scalarmag10}
\end{eqnarray}
The measure of the path integral \eqref{scalarmag9} is written as

\begin{equation}
\left[\EuScript{D}\mu\right]=\left[\EuScript{D}A\right]\left[\EuScript{D}b\right]\left[\EuScript{D}\bar{c}\right]\left[\EuScript{D}c\right]\left[\EuScript{D}\bar{\omega}\right]\left[\EuScript{D}\omega\right]\left[\EuScript{D}\bar{\varphi}\right]\left[\EuScript{D}\varphi\right]\left[\EuScript{D}\xi\right]\left[\EuScript{D}\tau\right]\left[\EuScript{D}\phi\right]\left[\EuScript{D}\bar{\eta}\right]\left[\EuScript{D}\eta\right]\left[\EuScript{D}\bar{\theta}\right]\left[\EuScript{D}\theta\right]\,.
\label{scalarmag11}
\end{equation}
At one-loop order, we should take the quadratic part of\footnote{For the moment, we can ignore the contribution from $S^{\mathrm{loc}}_{\mathrm{RGZ}}$, which is $(J,\tilde{J})$-independent.} $\Sigma (J,\tilde{J})$,

\begin{eqnarray}
\Sigma^{(2)}(J,\tilde{J}) &=& \int \dd^dx \left\{\frac{1}{2}\left[(\partial_{\mu}\phi^{\alpha})(\partial_{\mu}\phi^{\alpha})+(\partial_\mu \phi)(\partial_\mu \phi)\right]+\frac{m^2_\phi}{2}\phi^\alpha \phi^\alpha + \frac{m^2_\phi}{2}\phi\phi\right\}\nonumber\\
&-&\int \dd^dx\left(-\bar{\eta}^{\alpha\delta}\delta^{\alpha\beta}\partial^2\eta^{\beta\delta}+\bar{\theta}^{\alpha\delta}\delta^{\alpha\beta}\partial^2\theta^{\beta\delta}-g\sigma^2\epsilon^{\alpha\beta}\phi^{h,3}(\bar{\eta}+\eta)^{\alpha\beta}\right)\nonumber\\
&+& J\int \dd^dx~\phi^{h,3}\phi^{h,3}-\tilde{J}\int \dd^dx\left(\bar{\eta}^{\alpha\delta}\eta^{\alpha\beta}-\bar{\theta}^{\alpha\beta}\theta^{\alpha\beta}\right)\,.
\label{scalarmag12}
\end{eqnarray}
Integrating the auxiliary fields $(\tau^\alpha,\tau)$ which enforce the transversality condition of $A^{h,a}=(A^{h,\alpha}_\mu, A^h_\mu)$, we see that the gauge-invariant scalar field $\phi^{h,a}=(\phi^{h,\alpha}, \phi^{h,3})$ can be expressed as in eq.\eqref{fophi} with $\xi^a = \frac{\partial A^a}{\partial^2}$.

 Since we want to maintain the action $\Sigma^{(2)}$ to the quadratic order in the fields, we see that $\phi^{h,a}\approx \phi^a$. Hence, the $(J,\tilde{J})$-dependent part of the one-loop order vacuum energy $\mathcal{E}$ is given by

\begin{equation}
\mathcal{E}^{(1)}(J,\tilde{J}) = \frac{1}{2}\int \frac{\dd^dk}{(2\pi)^d}~\mathrm{ln}\left(k^2+m^2_\phi+2J+\frac{4g^2\sigma^4}{k^2+\tilde{J}}\right)\,.
\label{scalarmag13}
\end{equation}
This implies,

\begin{equation}
\langle \phi^{h,3}\phi^{h,3}\rangle_{\mathrm{1-loop}} = - \int \frac{\dd^dk}{(2\pi)^d}\frac{m^2_\phi}{k^4+m^2_\phi k^2+4g^2\sigma^4}-4g^2\sigma^4\int \frac{\dd^d k}{(2\pi)^d}\frac{1}{k^2}\frac{1}{k^4+m^2_\phi k^2 + 4g^2\sigma^4}\,,
\label{scalarmag14}
\end{equation}
and

\begin{equation}
\langle\bar{\eta}^{\alpha\beta}\eta^{\alpha\beta}-\bar{\theta}^{\alpha\beta}\theta^{\alpha\beta}\rangle_{\mathrm{1-loop}}=2g^2\sigma^4\int \frac{\dd^dk}{(2\pi)^d}\frac{1}{k^2}\frac{1}{k^4+m^2_\phi k^2+4g^2\sigma^4}\,.
\label{scalarmag15}
\end{equation}
Eq.\eqref{scalarmag15} shows that, at the one-loop level, the condensate of auxiliary fields, $\langle\bar{\eta}^{\alpha\beta}\eta^{\alpha\beta}-\bar{\theta}^{\alpha\beta}\theta^{\alpha\beta}\rangle_{\mathrm{1-loop}}$,  is ultraviolet convergent. For $d=3,4$, such a condensate is perfectly well-defined in the infrared region and can be safely introduced. In $d=2$, an infrared singularity at $k=0$ turns out to appear. This is in agreement with the refining condensates in the Gribov-Zwanziger setup. From eq.\eqref{scalarmag14}, we see that the condensate $\langle \phi^{h,3}\phi^{h,3}\rangle$ has two contributions: one proportional to $\tilde{m}^{2}_{\mathrm{diag}}$ which exists irrespective of the presence of $\sigma$ and the other one proportional to $\sigma^4$. The former contains an ultraviolet divergence which can be taken into account by the standard renormalization techniques while the latter is ultraviolet convergent and free from infrared divergences in $d=3,4$. In $d=2$, an infrared singularity appears preventing the introduction of this condensate. We must emphasize that this condensate does not affect the qualitative behavior of the initial theory.

%, but we should recognize that the different dynamics between the diagonal and off-diagonal sectors of the scalar field justifies the decomposition \eqref{scalarmag2}. 

Therefore, in $d=2$, the scalar field action non-perturbatively coupled with the gauge sector is given by \eqref{scalarmag5}, while in $d=3,4$ the condensates $\langle\bar{\eta}^{\alpha\beta}\eta^{\alpha\beta}-\bar{\theta}^{\alpha\beta}\theta^{\alpha\beta}\rangle$ and $\langle \phi^{h,3}\phi^{h,3}\rangle$ have to be taken into account, giving rise to the following refined action 

\begin{eqnarray}
S^\phi_{\mathrm{MAG-loc}} &=& S_\mathrm{scalar}-\int \dd^dx\left(\bar{\eta}^{\alpha\delta}\mathcal{M}^{\alpha\beta}(A^h)\eta^{\beta\delta}-\bar{\theta}^{\alpha\delta}\mathcal{M}^{\alpha\beta}(A^h)\theta^{\beta\delta}-g\sigma^2\epsilon^{\alpha\beta}\phi^{h,3}(\bar{\eta}+\eta)^{\alpha\beta}\right)\nonumber\\
&+& \frac{\mu^{2}_{\mathrm{diag}}}{2}\int \dd^dx~\phi^{h,3}\phi^{h,3}-\rho^2\int \dd^dx\left(\bar{\eta}^{\alpha\beta}\eta^{\alpha\beta}-\bar{\theta}^{\alpha\beta}\theta^{\alpha\beta}\right)\,.
\label{scalarmag16}
\end{eqnarray}
From the actions  \eqref{scalarmag16} and \eqref{scalarmag5} we can compute the tree-level Abelian component of the scalar field propagator. The expressions in $d=2$ and $d=3,4$ are, respectively,

\begin{equation}
\langle \phi(k) \phi(-k)\rangle_{d=2} = \frac{k^2}{k^4 + m^2_\phi k^2+4g^2\sigma^4}\,,
\label{scalarmag17}
\end{equation}
and 

\begin{equation}
\langle \phi(k) \phi(-k)\rangle_{d=3,4} = \frac{k^2+\rho^2}{k^4 + (m^2_\phi+\mu^2_{\mathrm{diag}}+\rho^2)k^2+(m^2_\phi+\mu^2_{\mathrm{diag}})\rho^2+4g^2\sigma^4}\,.
\label{scalarmag18}
\end{equation}
From eq.\eqref{scalarmag17} and \eqref{scalarmag18}, we see that the propagator of the Abelian component of the scalar field displays the same features observed for the tree-level propagator of the scalar field in the linear covariant and Curci-Ferrari gauges, eqs.\eqref{scalarlcg17},\eqref{scalarlcg17}.

%%%%%%%%%%%%%%%%%%%%%%%%%%%%%%%%%%%%%%%%%%%%%%%%%%%%%%%%%%%%%
\section{Generalization of the non-perturbative matter coupling for quark fields} \label{npspinor}
%%%%%%%%%%%%%%%%%%%%%%%%%%%%%%%%%%%%%%%%%%%%%%%%%%%%%%%%%%%%%

In the previous section, we have presented a prescription for the non-perturbative coupling of scalar fields in the adjoint representation of the gauge group with the gauge sector. Such a coupling arises from the introduction of a non-local term which  shares great similarity with the corresponding horizon term introduced in the gluon sector to implement the restriction of the domain of integration to the Gribov region. Interestingly, this term naturally appears  through the dimensional reduction of higher-dimensional Yang-Mills theory \cite{Guimaraes:2016okb}.

In the present section we follow the same reasoning for the case of fermionic matter fields in the fundamental representation of the gauge group. This case is particularly important since it allows us to obtain an analytic non-perturbative expression of the quark field propagator. As before, we divide the analysis in two subsections for linear covariant/Curci-Ferrari gauges and for the maximal Abelian gauge. We have collected our conventions regarding spinors and related issues in Appendix~A.

%%%%%%%%%%%%%%%%%%%%%%%%%%%%%%%%%%%%%%%%%%%%%%%%%%%%%%%%%%%%%
\subsection{Linear covariant and Curci-Ferrari gauges} \label{spinorLCGCF}
%%%%%%%%%%%%%%%%%%%%%%%%%%%%%%%%%%%%%%%%%%%%%%%%%%%%%%%%%%%%%

Let us begin by considering the Dirac action in Euclidean space minimally coupled with the gauge sector,

\begin{equation}
S_{\mathrm{Dirac}} = \int \dd^dx\left[\bar{\psi}^{I}\gamma_\mu D^{IJ}_{\mu}\psi^{J}-m_\psi \bar{\psi}^{I} \psi^{I}\right]\,,
\label{spinorlcg1}
\end{equation}
where capital latin indices $\left\{I,J,\ldots\right\}$ stand for the fundamental representation of $SU(N)$. The covariant derivative $D^{IJ}_{\mu}$ is defined by

\begin{equation}
D^{IJ}_\mu = \delta^{IJ}\partial_\mu - ig(T^a)^{IJ}A^{a}_\mu\,,
\label{spinorlcg2}
\end{equation} 
with $T^a$ the generators of $SU(N)$ in the fundamental representation. In strict analogy to what has been proposed in Sect.~\ref{noscalaradjoint}, the non-perturbative fermion matter coupling is introduced  by adding to the Dirac action the non-local term  
\begin{equation}
\EuScript{H}(\psi^h) = -g^2\int \dd^dx \dd^dy~\bar{\psi}^{h,I}(x)(T^a)^{IJ}\left[\EuScript{M}^{-1}(A^h)\right]^{ab}(x,y)(T^b)^{JK}\psi^{h,K}(y)\,,
\label{spinorlcg3}
\end{equation}
with  $\left( \EuScript{M}(A^h) \right)^{ad}$ given by eq.\eqref{npbrst6} and where the gauge-invariant spinor $\psi^h$ is defined as

\begin{equation}
\psi^{h,I} = \psi^{I} - ig\frac{1}{\partial^2}(\partial_\mu A^a_\mu)(T^a)^{IJ}\psi^{J}+\mathcal{O}(A^2)\,.
\label{spinorlcg4}
\end{equation}
Employing the Stueckelberg field $\xi^a$, the all order BRST invariant spinor field  $\psi^h$ is obtained as  

\begin{equation}
\psi^{h} = h^\dagger \psi = \mathrm{e}^{-ig\xi^a T^a}\psi\,.
\label{spinorlcg5}
\end{equation}
From 
\begin{equation} 
s h^{\dagger} = ig h^{\dagger} c \;, \qquad s \psi = -ig c \psi \;, \label{hp}
\end{equation}
it immediately follows that  $\psi^{h}$ is BRST invariant, namely 
\begin{equation}
s \psi^h = 0 \;. \label{inph} 
\end{equation}
Solving the transversality condition $\partial_\mu A^{h,a}_\mu = 0$ for the Stueckelberg field $\xi^a$ and plugging it in eq.\eqref{spinorlcg5}, see Appendix~A of \cite{Capri:2015ixa},  we reobtain expression \eqref{spinorlcg4}. Hence, following the prescription discussed in \cite{Capri:2014bsa,Capri:2016aqq}, the fermionic action non-perturbatively coupled to the gauge sector is given by

\begin{equation}
S^{\psi} = S_{\mathrm{Dirac}}+M^3\EuScript{H}(\psi^h)\,,
\label{spinorlcg6}
\end{equation}
where $M$ is the analogue of the Gribov parameter $\gamma$ for the fermionic sector. 

The term $\EuScript{H}(\psi^h)$ is non-local due to the inverse of $\EuScript{M}(A^h)$, eq.\eqref{npbrst6}. Nevertheless, the action \eqref{spinorlcg6} can be localized in complete analogy with the localization of the Gribov-Zwanziger action by means of the introduction of commuting spinor fields $(\bar{\theta},\theta)^{aI}$ as well as of anti-commuting ones $(\bar{\lambda},\lambda)^{aI}$. The local form of expression \eqref{spinorlcg3} is given by

\begin{equation}
M^3\EuScript{H}(\psi^h)\,\longrightarrow\,\int \dd^dx\left(\bar{\theta}^{aI}\EuScript{M}^{ab}(A^h)\theta^{bI}-\bar{\lambda}^{aI}\EuScript{M}^{ab}(A^h)\lambda^{bI}-gM^{3/2}\bar{\lambda}^{aI}(T^a)^{IJ}\psi^{h,J}+gM^{3/2}\bar{\psi}^{h,I}(T^a)^{IJ}\lambda^{aJ}\right)\,,
\label{spinorlcg7}
\end{equation}
which, upon integration over the auxiliary fields $(\bar{\theta},\theta)^{aI}$ and $(\bar{\lambda},\lambda)^{aI}$, gives back the non-local quantity of  eq.\eqref{spinorlcg3}

Therefore, the local action with the non-perturbative coupling between fermionic matter and the gauge sector is expressed as

\begin{equation}
\tilde{S}^{\psi}_{\mathrm{loc}} = S_{\mathrm{Dirac}} + \int \dd^dx\left(\bar{\theta}^{aI}\EuScript{M}^{ab}(A^h)\theta^{bI}-\bar{\lambda}^{aI}\EuScript{M}^{ab}(A^h)\lambda^{bI}-gM^{3/2}\bar{\lambda}^{aI}(T^a)^{IJ}\psi^{h,J}+gM^{3/2}\bar{\psi}^{h,I}(T^a)^{IJ}\lambda^{aJ}\right)\,.
\label{spinorlcg8}
\end{equation}
As extensively discussed in the present work, the presence of the parameter $M$, akin to the Gribov parameter $\gamma$, and of the quadratic coupling between the auxiliary localizing fields and the corresponding matter field give rise to a dynamical and non-perturbative instability, resulting in the formation of condensates. Again, we present the one-loop computation which hints the existence of such condensates. To do so, we introduce the following operators 
\begin{equation}
-J\int \dd^dx~\bar{\psi}^{h,I}\psi^{h,I}\,\,\,\,\,\,\mathrm{and}\,\,\,\,\,\,\tilde{J}\int \dd^dx\left(\bar{\theta}^{aI}\theta^{aI}-\bar{\lambda}^{aI}\lambda^{aI}\right)\,,
\label{spinorlcg9}
\end{equation}
into the action \eqref{spinorlcg8},  yielding 

\begin{equation}
\Sigma (J,\tilde{J}) = \tilde{S}^{\psi}_\mathrm{loc}-J\int \dd^dx~\bar{\psi}^{h,I}\psi^{h,I}+\tilde{J}\int \dd^dx\left(\bar{\theta}^{aI}\theta^{aI}-\bar{\lambda}^{aI}\lambda^{aI}\right)\,.
\label{spinorlcg10}
\end{equation}
We aim at computing the following  condensates:

\begin{equation}
\langle \bar{\psi}^{h,I}(x)\psi^{h,I}(x)\rangle\,\,\,\,\,\,\mathrm{and}\,\,\,\,\,\,\langle \bar{\theta}^{aI}(x)\theta^{aI}(x)-\bar{\lambda}^{aI}(x)\lambda^{aI}(x)\rangle\,,
\label{spinorlcg11}
\end{equation}
which can be obtained by taking taking the derivatives with respect to $(J,\tilde{J})$ of the vacuum energy\footnote{We restrict ourselves to the contributions relevant for our purposes.} $\mathcal{E}(J,\tilde{J})$ at one-loop order,

\begin{equation}
\mathrm{e}^{-V\mathcal{E}^{(1)}} = \int \left[\EuScript{D}\mu\right]\mathrm{e}^{-\Sigma^{(2)}(J,\tilde{J})}\,,
\label{spinorlcg12}
\end{equation}
with $\Sigma^{(2)}(J,\tilde{J})$ the quadratic part of $\Sigma(J,\tilde{J})$, namely

\begin{eqnarray}
\langle \bar{\psi}^{h,I}(x)\psi^{h,I}(x)\rangle &=& -\frac{\partial\mathcal{E}^{(1)}}{\partial J}\Big|_{J=\tilde{J}=0}\,,\nonumber\\
\langle \bar{\theta}^{aI}(x)\theta^{aI}(x)-\bar{\lambda}^{aI}(x)\lambda^{aI}(x)\rangle &=& \frac{\partial\mathcal{E}^{(1)}}{\partial \tilde{J}}\Big|_{J=\tilde{J}=0}\,.
\label{spinorlcg13}
\end{eqnarray}

Explicitly, $\Sigma^{(2)}(J,\tilde{J})$ is written as

\begin{eqnarray}
\Sigma^{(2)}(J,\tilde{J}) &=& \int \dd^dx\left[\bar{\psi}^{I}\gamma_\mu \partial_\mu\psi^I-m_\psi\bar{\psi}^I\psi^I+\bar{\lambda}^{aI}\partial^2\lambda^{aI}-\bar{\theta}^{aI}\partial^2\theta^{aI}-gM^{3/2}\bar{\lambda}^{aI}(T^a)^{IJ}\psi^{J}+gM^{3/2}\bar{\psi}^{I}(T^a)^{IJ}\lambda^{aJ}\right.\nonumber\\
&-&\left.J\bar{\psi}^{I}\psi^{I}+\tilde{J}(\bar{\theta}^{aI}\theta^{aI}-\bar{\lambda}^{aI}\lambda^{aI})\right]\,.
\label{spinorlcg14}
\end{eqnarray}
Performing the path integral over the auxiliary localizing fields yields the following expression
\begin{equation}
\int \left[\EuScript{D}\bar{\psi}\right]\left[\EuScript{D}\psi\right]\mathrm{exp}\left\{\int \frac{\dd^dk}{(2\pi)^d}\bar{\psi}^{I}(k)\left[\delta^{IJ}\gamma_\mu(ik_\mu)+\delta^{IJ}(m_\psi+J)+g^2M^3\frac{(T^a)^{IK}(T^a)^{KJ}}{k^2+\tilde{J}}\right]\psi^{J}(-k)\right\}\,.
\label{spinorlcg15}
\end{equation}
Making use of the relation

\begin{equation}
(T^a)^{IK}(T^a)^{KJ}=\delta^{IJ}\frac{N^2-1}{2N}\,,
\label{spinorlcg16}
\end{equation}
and performing the path integral over $(\bar{\psi},\psi)$, one obtains

\begin{equation}
\mathrm{det}\left\{\delta^{IJ}\left[\gamma_\mu (ik_\mu)+\left(m_\psi + J+g^2M^3\frac{N^2-1}{2N}\frac{1}{p^2+\tilde{J}}\right)\mathds{1}\right]\right\}\,.
\label{spinorlcg17}
\end{equation}
After simple manipulations and employing the identity

\begin{equation}
\mathrm{det}(i\gamma_\mu k_\mu+A\mathds{1}) = \mathrm{det}^{1/2}\left(k^2\mathds{1}+A^2\mathds{1}\right)\,,
\label{spinorlcg18}
\end{equation}
one ends up with

\begin{equation}
\mathrm{e}^{-V\mathcal{E}^{(1)}}=\left\{\mathrm{det}\left[k^2\mathds{1}+\left(m_\psi+J+g^2M^3\frac{N^2-1}{2N}\frac{1}{k^2+\tilde{J}}\right)^2\mathds{1}\right]\right\}^{(N^2-1)/2}\,.
\label{spinorlcg19}
\end{equation}

From \eqref{spinorlcg19} it is immediate to extract the vacuum energy $\mathcal{E}^{(1)}$, which is written as

\begin{equation}
\mathcal{E}^{(1)}(J,\tilde{J})=-2(N^2-1)\int\frac{\dd^dk}{(2\pi)^d}\mathrm{ln}\left[k^2+\left(m_\psi+J+g^2M^3\frac{N^2-1}{2N}\frac{1}{k^2+\tilde{J}}\right)^2\right]\,.
\label{spinorlcg20}
\end{equation}
Finally, we are ready to compute the expectation values \eqref{spinorlcg13}, by differentiating \eqref{spinorlcg20} with respect to the sources $(J,\tilde{J})$ as in \eqref{spinorlcg13}. One obtains,

\begin{eqnarray}
\langle \bar{\psi}^{I}\psi^I\rangle_{\mathrm{1-loop}} &=& 4(N^2-1)m_\psi\int \frac{\dd^d k}{(2\pi)^d}\frac{k^4}{k^6+\left(m_\psi k^2+g^2M^3\frac{N^2-1}{2N}\right)^2}\nonumber\\
&-&g^2M^3\frac{(N^2-1)^2}{N}\int \frac{\dd^d k}{(2\pi)^d}\frac{m^2_\psi+\left(g^2M^3\frac{N^2-1}{2N}\right)^2\frac{1}{k^4}+g^2M^3m_\psi\frac{N^2-1}{n}\frac{1}{k^2}}{k^6+\left(m_\psi k^2+g^2M^3\frac{N^2-1}{2N}\right)^2}\,,
\label{spinorlcg21}
\end{eqnarray}
and

\begin{equation}
\langle \bar{\theta}^{aI}\theta^{aI}-\bar{\lambda}^{aI}\lambda^{aI}\rangle_{\mathrm{1-loop}} = 2g^2M^3\frac{(N^2-1)^2}{N}\int \frac{\dd^dk}{(2\pi)^d}\frac{m_\psi+g^2M^3\frac{N^2-1}{2N}\frac{1}{k^2}}{k^6+\left(m_\psi k^2+g^2M^3\frac{N^2-1}{2N}\right)^2}\,,
\label{spinorlcg22}
\end{equation}
where the prescriptions of the dimensional regularization were employed. In $d=4$, we see from eq.\eqref{spinorlcg21} that the contribution which is directly proportional to the parameter $M$ is perfectly ultraviolet convergent. This is in agreement with the fact that the introduction of the non-local term of the type of eq.\eqref{spinorlcg3} does not introduce any new ultraviolet divergence \cite{Capri:2015mna}. From eq.\eqref{spinorlcg22}, we easily see that the one-loop contribution to the condensate $\langle \bar{\theta}^{aI}\theta^{aI}-\bar{\lambda}^{aI}\lambda^{aI}\rangle$ is non-vanishing and ultraviolet convergent. These results show explicitly, already at one-loop order,  that the introduction of the non-perturbative matter coupling \eqref{spinorlcg22}  contributes definitively to the formation of such condensates. 

As usual, the dynamical formation of those condensates can be taken into account from the beginning by refining the matter sector in the following way

\begin{equation}
\tilde{S}^{\psi}_{\mathrm{loc}}\,\longrightarrow\, {S}^{\psi}_{\mathrm{loc}} = \tilde{S}^{\psi}_{\mathrm{loc}} -\tilde{m}_{\psi}\int \dd^dx~\psi^{h,I}\psi^{h,I} + \rho^2\int \dd^dx\left( \bar{\theta}^{aI}\theta^{aI}-\bar{\lambda}^{aI}\lambda^{aI} \right)\,.
\label{spinorlcg23}
\end{equation}
Finally, one can compute the quark field propagator at tree level from the refined action \eqref{spinorlcg23}. The result is

\begin{equation}
\langle \bar{\psi}^{I}(-p)\psi^{J}(p)\rangle = - \delta^{IJ}\frac{-i\gamma_\mu p_\mu+\left(M_\psi+g^2M^3\frac{(N^2-1)}{2N}\frac{1}{p^2+\rho^2}\right)}{p^2+\left(M_\psi+g^2M^3\frac{(N^2-1)}{2N}\frac{1}{p^2+\rho^2}\right)^2}\,,
\label{spinorlcg24}
\end{equation}
with $M_\psi = m_\psi + \tilde{m}_\psi$.

The propagator \eqref{spinorlcg24} is the same as the one computed in the Landau gauge \cite{Capri:2014bsa}, {\it i.e.}  $\alpha=0$.  Of course, higher orders correction will, eventually, introduce some $\alpha$-dependence in \eqref{spinorlcg24}. In the particular case of $\alpha = 0$, the propagator \eqref{spinorlcg24} fits well recent lattice data, see \cite{Capri:2014bsa} and references therein. To the best of our knowledge, there are no available  numerical simulations of the quark propagator in linear covariant gauges. Hence, our result could be a motivation for such an endeavour in the near future.

As described in the case of scalar fields, the generalization of the present construction to the case of the Curci-Ferrari gauges is straightforward. In particular, the results  obtained here also hold in the Curci-Ferrari gauge, which differs from the linear covariant gauges by  non-linear terms which do not contribute to the order we are dealing with. In particular, the quark propagator at the tree-level remains the same as in eq.\eqref{spinorlcg24}.

%%%%%%%%%%%%%%%%%%%%%%%%%%%%%%%%%%%%%%%%%%%%%%%%%%%%%%%%%%%%%
\subsection{Maximal Abelian gauge}
%%%%%%%%%%%%%%%%%%%%%%%%%%%%%%%%%%%%%%%%%%%%%%%%%%%%%%%%%%%%%

In this subsection, we proceed with the analysis of the non-perturbative coupling of quark matter fields in the maximal Abelian gauge case. In full analogy with the case of the scalar matter field, eq.\eqref{scalarmag3}, for the non-perturbative BRST invariant coupling in  the quark sector we write

\begin{equation}
\EuScript{H}_{\mathrm{MAG}}(\psi^h)=-g^2\int \dd^dx \dd^dy~\bar{\psi}^{h,I}(x)(T^\alpha)^{IJ}\left[\mathcal{M}^{-1}(A^h)\right]^{\alpha\beta}(x,y)(T^\beta)^{JK}\psi^{h,K}(y)\,,
\label{spinormag1}
\end{equation}
where the gauge-invariant field $\psi^h$ is defined by eq.\eqref{spinorlcg5}, while the Faddeev-Popov operator in the maximal Abelian gauge, $\mathcal{M}^{\alpha\beta}$, is given by eq.\eqref{mag8}. The non-perturbative coupling of quark matter fields  with the gauge sector in the maximal Abelian gauge is thus given by

\begin{equation}
S^{\psi} = S_{\mathrm{Dirac}} + M^3\EuScript{H}_{\mathrm{MAG}}(\psi^h)\,,
\label{spinormag2}
\end{equation}
where, as before, the parameter $M$ plays an analogue role of the Gribov parameter $\gamma$ in the matter sector. As exhaustively discussed in the previous sections, the non-local  quark matter term \eqref{spinormag1} can be localized by means of auxiliary fields. The gauge-invariant field $\psi^h$ can be written in local form in the same manner described in Subsect.~\ref{spinorLCGCF}. On the other hand, a pair of commuting $(\bar{\theta},\theta)^{\alpha I}$ and anticommuting $(\bar{\lambda},\lambda)^{\alpha I}$ fields are introduced in order to localize $\EuScript{H}_{\mathrm{MAG}}(\psi^h)$, namely,

\begin{equation}
M^3\EuScript{H}_{\mathrm{MAG}}(\psi^h)\,\longrightarrow\, \int \dd^dx\left(\bar{\theta}^{\alpha I}\mathcal{M}^{\alpha\beta}(A^h)\theta^{\beta I}-\bar{\lambda}^{\alpha I}\mathcal{M}^{\alpha \beta}(A^h)\lambda^{\beta I}-gM^{3/2}\bar{\lambda}^{\alpha I}(T^{\alpha})^{IJ}\psi^{h,J}+gM^{3/2}\bar{\psi}^{h,I}(T^\alpha)^{IJ}\lambda^{\alpha J}\right)\,.
\label{spinormag3}
\end{equation}
Therefore, the action of  quark matter fields coupled with the gauge sector in a non-perturbative way is expressed, in local form, as

\begin{equation}
\tilde{S}^{\psi}_{\mathrm{MAG-loc}} = S_{\mathrm{Dirac}} + \int \dd^dx\left(\bar{\theta}^{\alpha I}\mathcal{M}^{\alpha\beta}(A^h)\theta^{\beta I}-\bar{\lambda}^{\alpha I}\mathcal{M}^{\alpha \beta}(A^h)\lambda^{\beta I}-gM^{3/2}\bar{\lambda}^{\alpha I}(T^{\alpha})^{IJ}\psi^{h,J}+gM^{3/2}\bar{\psi}^{h,I}(T^\alpha)^{IJ}\lambda^{\alpha J}\right)\,.
\label{spinormag4}
\end{equation}
At this stage, it is not unexpected to predict that, again, the  action \eqref{spinormag4} suffers from dynamical non-perturbative  instabilities, giving rise to the formation of condensates. The procedure to explicit check the existence of such codensates goes exactly along the same lines of the previous case, namely: constant sources $J$ and $\tilde{J}$ are coupled to the composite operators $\bar{\psi}^{h,I}\psi^{h,I}$ and $(\bar{\theta}^{\alpha I}\theta^{\alpha I}-\bar{\lambda}^{\alpha I}\lambda^{\alpha I})$, \textit{i.e.}

\begin{equation}
-J\int \dd^dx~\bar{\psi}^{h,I}\psi^{h,I}\,\,\,\,\,\,\mathrm{and}\,\,\,\,\,\,\tilde{J}\int\dd^dx\left(\bar{\theta}^{\alpha I}\theta^{\alpha I}-\bar{\lambda}^{\alpha I}\lambda^{\alpha I}\right)\,,
\label{spinormag5}
\end{equation}
which are introduced in the action \eqref{spinormag4},  giving rise to 

\begin{equation}
\Sigma (J,\tilde{J}) = \tilde{S}^{\psi}_{\mathrm{MAG-loc}} -J\int \dd^dx~\bar{\psi}^{h,I}\psi^{h,I}+\tilde{J}\int\dd^dx\left(\bar{\theta}^{\alpha I}\theta^{\alpha I}-\bar{\lambda}^{\alpha I}\lambda^{\alpha I}\right)\,.
\label{spinormag6}
\end{equation}
The condensates are obtained by taking derivatives of the vacuum energy $ \mathcal{E} $ corresponding to the action \eqref{spinormag6} with respect to the sources $J$ and $\tilde{J}$, and setting them to zero, \textit{i.e.}

\begin{eqnarray}
\langle \bar{\psi}^{h,I}(x){\psi}^{h,I}(x)\rangle &=& - \frac{\partial \mathcal{E}}{\partial J}\Big|_{J = \tilde{J} = 0}\,,\nonumber\\ 
\langle \bar{\theta}^{\alpha I}(x)\theta^{\alpha I}(x)-\bar{\lambda}^{\alpha I}(x)\lambda^{\alpha I}(x) \rangle &=&  \frac{\partial \mathcal{E}}{\partial \tilde{J}}\Big|_{J = \tilde{J} = 0}\,,
\label{spinormag7}
\end{eqnarray}
with

\begin{equation}
\mathrm{e}^{-V\mathcal{E}} = \int \left[\EuScript{D}\bar{\psi}\right]\left[\EuScript{D}\psi\right]\left[\EuScript{D}\mu\right]\mathrm{e}^{-\Sigma (J,\tilde{J})}\,.
\label{spinormag8}
\end{equation}
At one-loop order, using the same techniques presented in Sect.~\ref{noscalaradjoint} and Subsect.~\ref{spinorLCGCF}, one obtains

\begin{equation}
\mathcal{E}^{(1)}(J,\tilde{J}) = -4 \int \frac{\dd^d k}{(2\pi)^d}\mathrm{ln}\left[k^2+\left(m_\psi+J+\frac{g^2 M^3}{2}\frac{1}{k^2+\tilde{J}}\right)^2\right]\,.
\label{spinormag9}
\end{equation}
Plugging eq.\eqref{spinormag9} into eq.\eqref{spinormag7}, one immediately gets

\begin{equation}
\langle \bar{\psi}^{h,I}(x) {\psi}^{h,I}(x)\rangle = 8\int \frac{\dd^dk}{(2\pi)^d}\frac{m_\psi k^4}{k^6+(m_\psi k^2 + \frac{g^2 M^3}{2})^2}-g^2 M^3 \int \frac{\dd^d k}{(2\pi)^d}\frac{\frac{g^2 m_\psi M^2}{k^2}+\frac{g^4 M^6}{4k^4}}{k^6+(m_\psi k^2 + \frac{g^2 M^3}{2})^2}
\label{spinormag10}
\end{equation} 
and 

\begin{equation}
\langle \bar{\theta}^{\alpha I}(x)\theta^{\alpha I}(x)-\bar{\lambda}^{\alpha I}(x)\lambda^{\alpha I}(x) \rangle = 4g^2 M^3 \int \frac{\dd^dk}{(2\pi)^d}\frac{m_\psi+\frac{g^2 M^3}{2}\frac{1}{k^2}}{k^6+\left(m_\psi k^2 + \frac{g^2 M^3}{2}\right)^2}\,.
\label{spinormag11}
\end{equation}

Once again, one notices that the contributions proportional to $M$ are ultraviolet finite. As such, we find already at one-loop order that such condensates are non-vanishing, due to the introduction of the non-perturbative coupling \eqref{spinormag1} in the quark matter sector. We should emphasize that, unlike the case of the linear covariant and Curci-Ferrari gauges, the condensate of the auxiliary fields, $ \langle \bar{\theta}^{\alpha I}(x)\theta^{\alpha I}(x)-\bar{\lambda}^{\alpha I}(x)\lambda^{\alpha I}(x) \rangle$, is purely diagonal. This is a direct consequence of the decomposition into diagonal and off-diagonal indices of the maximal Abelian gauge. Finally, as before, the dynamical generation of the condensates \eqref{spinormag7} can be taken into account by the refinement of the quark action, {\it i.e.} 

\begin{equation}
\tilde{S}^{\psi}_{\mathrm{MAG-loc}}\,\longrightarrow\, S^{\psi}_{\mathrm{MAG-loc}} = \tilde{S}^{\psi}_{\mathrm{MAG-loc}} - \tilde{m}_{\psi}\int \dd^dx \bar{\psi}^{h,I}\psi^{h,I}+\rho^2\int \dd^dx\left(\bar{\theta}^{\alpha I}\theta^{\alpha I}-\bar{\lambda}^{\alpha I}\lambda^{\alpha I} \right)\,.
\label{spinormag12}
\end{equation}
Out of the action \eqref{spinormag12}, one can compute the tree-level quark propagator, which is given by

\begin{equation}
\langle \bar{\psi}^{I}(-k) \psi^{J}(k)\rangle = -\delta^{IJ}\frac{-i\gamma_\mu k_\mu+\left(M_\psi+\frac{g^2M^3}{2}\frac{1}{k^2+\rho^2}\right)}{k^2+\left(M_\psi+\frac{g^2M^3}{2}\frac{1}{k^2+\rho^2}\right)^2}\,.
\label{spinormag13}
\end{equation}
Quite importantly, the tree-level quark propagator \eqref{spinormag13} is in qualitative agreement with the very recent lattice results reported in \cite{Schrock:2015pna}. Such an agreement works as a highly non-trivial check of the non-perturbative matter coupling proposed here.

%%%%%%%%%%%%%%%%%%%%%%%%%%%%%%%%%%%%%%%%%%%%%%%%%%%%%%%%%%%%%%
\section{Conclusions}
%%%%%%%%%%%%%%%%%%%%%%%%%%%%%%%%%%%%%%%%%%%%%%%%%%%%%%%%%%%%%%

In this work, we have extended the non-perturbative gauge-matter coupling proposed in \cite{Capri:2016aqq,Capri:2014bsa} to linear covariant, Curci-Ferrari and maximal Abelian gauges. In particular, we have investigated the coupling of scalar fields in the adjoint representation of the gauge group as well as of quark fields in the fundamental representation.  

The non-perturbative nature of the proposal relies on the introduction of an additional term in which the matter fields are coupled to the inverse of the operator $\EuScript{M}(A^h)$, whose existence is ensured by the restriction of the domain of integration in the functional integral to the Gribov region. As discussed in details throughout the paper,  this additional term in the matter fields shares great similarity with the horizon function  introduced in the pure gauge sector in order to implement the restriction to the Gribov region. Albeit non-local, the resulting action can be cast in local form by the introduction of auxiliary fields which, as in the case of the localizing Zwanziger fields of the pure gauge sector, develop their own dynamics giving rise to the formation of condensates, as  explicitly checked through one-loop computations. Moreover, the condensates arising in the matter sector can be taken into account through an effective action which looks much alike the Refined Gribov-Zwanziger action which accounts for the existence of similar condensates in the gluon sector.  Out of this action, the tree-level propagators for matter fields were analysed, giving rise to reflection positivity violating propagators. As in the case of the gluon propagator, the positivity violation is taken as a signal that colored matters fields are confined too. 

We emphasize that the final effective action which encodes the non-perturbative effects of the matter sector is invariant under BRST transformations. This was achieved by the introduction of the suitable gauge-invariant fields $A^h$, $\phi^h$ and $\psi^h$,  see \cite{Capri:2015ixa,Capri:2016aqq,Capri:2015nzw,Pereira:2016fpn,Capri:2016gut}, which, albeit local,  are non-polynomial in the auxiliary Stueckelberg type field $\xi^a$. Nevertheless, such variables as well as the proposed non-perturbative matter coupling give rise to a local ation which can be proven to be renormalizable to all orders,  see \cite{Fiorentini:2016rwx,Capri:2017}.

The present work can give rise to several future investigations among which we quote: \textit{i)} as done in the pure gauge sector  \cite{Capri:2016gut}, we are now ready for a detailed  analysis of the Nielsen identities, in the case of linear covariant gauges, to investigate the independence of the poles of the matter field propagator from the gauge parameter $\alpha$; \textit{ii)} use the gauge-invariance of $A^h$, $\phi^h$ and $\psi^h$ to explore the Landau-Khalatnikov-Fradkin tranformations, as briefly discussed in \cite{Capri:2016gut} for the gluon sector, and analyse how gauge-matter correlators depend on the gauge parameter $\alpha$, while checking out how the results compare with those obtained through the aforementioned Nielsen identitites; \textit{iii)} study of how the presence of the Higgs mechanism can drive the transition between the confining and de-confining regimes in a BRST invariant fashion, \textit{iv)} investigate how the present proposal generalizes to supersymmetric gauge theories,  \textit{v)} stimulate different groups from other approaches such as lattice simulations and Dyson-Schwinger equations to study two-point functions of matter fields away from Landau gauge. As in the gluon sector, the interplay between different approaches in the study of non-perturbative correlation functions will certainly be very successful.

%%%%%%%%%%%%%%%%%%%%%%%%%%%%%%%%%%%%%%%%%%%%%%%%%%%%%%%%%%%%%%
\section*{Acknowledgements}
%%%%%%%%%%%%%%%%%%%%%%%%%%%%%%%%%%%%%%%%%%%%%%%%%%%%%%%%%%%%%%

The Conselho Nacional de Desenvolvimento Cient\'{i}fico e Tecnol\'{o}gico (CNPq-Brazil) and The Coordena\c c\~ao de Aperfei\c coamento de Pessoal de N\'ivel Superior (CAPES) are acknowledged.

\appendix

\section{Conventions in Euclidean space}

The gamma matrices $\gamma_\mu$ obey the Clifford algebra

\begin{equation}
\left\{\gamma_\mu,\gamma_\nu\right\}=2\delta_{\mu\nu}\,,
\label{apb1}
\end{equation}
with
\begin{equation}
\gamma_4 = {\begin{pmatrix}
0 & \mathds{1} \\ 
\mathds{1} & 0 
\end{pmatrix}}\,,\,\,\,\,\,\gamma_k=-i{\begin{pmatrix}
0 & \sigma_k \\ 
-\sigma_k  & 0 
\end{pmatrix}}\,,
\label{apb2}
\end{equation}
and

\begin{equation}
\sigma_4={\begin{pmatrix}
 1& 0 \\ 
 0& 1
\end{pmatrix}}\,,\,\,\,\,\sigma_1={\begin{pmatrix}
0 & 1\\ 
1 & 0
\end{pmatrix}}\,,\,\,\,\,\sigma_2={\begin{pmatrix}
0 & -i\\ 
i &  0
\end{pmatrix}}\,,\,\,\,\,\sigma_3={\begin{pmatrix}
1 & 0\\ 
0 & -1
\end{pmatrix}}\,.
\label{apb3}
\end{equation}

%%%%%%%%%%%%%%%%%%%%%%%%%%%%%%%%%%%%%%%%%%%%%%%%%%%%%%%%%%%%%%

%%%%%%%%%%%%%%%%%%%%%%%%%%%%%%

\end{document}